\documentclass[12pt]{article}

\title{The infrared problem in QED: \\ 
A lesson from a model with Coulomb interaction and realistic photon emission}

\author{ G. Morchio\footnote{e-mail: morchio@df.unipi.it, tel.:
0039 050 2214929, fax: 0039 050 2214887} \  
and F. Strocchi \\ INFN, Sezione di Pisa, Italy}
\date{}

\makeglossary

\newtheorem{Theorem}{Theorem}[section]

\newtheorem{Proposition}[Theorem]{Proposition}
\newtheorem{Lemma}[Theorem]{Lemma}

\usepackage{latexsym}
\usepackage[T1]{fontenc}
\usepackage{amsmath}
\usepackage{amssymb}

\def \ra {\rightarrow}

\def \endproof {\hfill \ensuremath{\Box}}
\def \AO {{\cal A}({\cal O})}
\def \AO' {{\cal A}({\cal O}')}

\def \Pf {{\bf Proof.\,\,}}

\def \be {\begin{equation}}
\def \ee {\end{equation}}
\def \ume {{\scriptstyle{\frac{1}{2}}}}
\def \ra {\rightarrow}

\def \eqq {\equiv}

\def \a {{\alpha}}
\def \b {{\beta}}

\def \d {{\delta}}
\def \eps {{\varepsilon}}

\def \l {{\lambda}}

\def \om {{\omega}}

\def \A {{\cal A}}

\def \D {{\cal D}}
\def \F {{\cal F}}
\def \G {{\cal G}}
\def \H {\mbox{${\cal H}$}}

\def \P {{\cal P}}
\def \S {{\cal S}}

\def \U {{\cal U}}
\def \V {{\cal V}}
\def \W {{\cal W}}

\def \id {{\bf 1 }}

\def \d^nu {{\partial^\nu}}
\def \d^la {{\partial^\lambda}}
\def \d^o {{\partial^0}}

\def \kbf {{\bf k}}

\def \q {{\bf q}}

\def \p {{\bf p}}

\def \x {{\bf x}}

\def\doppio#1{{\rm I}\kern-.1667em{\rm #1}}

\def\dR{{\rm \doppio R}}
\def\Q{\text{Q}\kern-.52em
    \text{\vrule height1.5ex width.5pt depth0pt}\kern.45em}
\def \Rbf {{\dR}}

\def\dZ{{\mathchoice {\hbox{$\Ss\textstyle Z\kern-0.4em Z$}}
{\hbox{$\Ss\textstyle Z\kern-0.4em Z$}} {\hbox{$\Ss\scriptstyle
Z\kern-0.25em Z$}} {\hbox{$\Ss\scriptscriptstyle Z\kern-0.2em
Z$}}}}

\def\dC{{\mathchoice{\hbox{$\rm\textstyle\text{\kern.35em\vrule
   height1.5ex width.5pt depth0pt\kern-.35em C}$}}
{\hbox{$\rm\textstyle\text{\kern.35em\vrule
   height1.5ex width.5pt depth0pt\kern-.35em C}$}}
{\hbox{$\rm\scriptstyle\text{\kern.35em\vrule
   height1.5ex width.3pt depth0pt\kern-.35em C}$}}
{\hbox{$\rm\scriptscriptstyle\text{\kern.35em\vrule
   height1.5ex width.2pt depth0pt\kern-.35em C}$}}}}

\makeatletter

\@addtoreset{equation}{section}

\begin{document}

\maketitle

\begin{abstract}
The scattering of photons and
heavy classical Coulomb interacting particles, 
with realistic particle-photon interaction (without particle recoil)
is studied adopting the Koopman formulation for the particles.
The model is translation invariant and
allows for a complete control of the Dollard strategy 
devised by Kulish-Faddeev and Rohrlich (KFR) for QED: in the adiabatic 
formulation, the M{\o}ller operators exist as strong limits 
and interpolate between the dynamics and a non-free  
asymptotic dynamics, which is a unitary group; the $S$-matrix
is non-trivial and exhibits the factorization of all the infrared divergences.
The implications of the KFR strategy on the open questions
of the LSZ asymptotic limits 
in QED are derived in the field theory version of 
the model, with the charged particles described by second quantized fields:
i) asymptotic limits of the charged fields, 
$\Psi_{out/in}(x)$,  are obtained as strong limits of modified 
LSZ formulas, with corrections given by a Coulomb 
phase operator and an exponential of the photon field; 
ii) free asymptotic electromagnetic fields, $B_{out/in}(x)$, 
are given by the massless LSZ formula, as in Buchholz approach;
\, iii) the asymptotic field algebras are a semidirect product
of the canonical algebras generated by  $B_{out/in}$, $\Psi_{out/in}$; 
\, iv)
on the asymptotic spaces, the Hamiltonian is the sum of the free 
(commuting) Hamiltonians of $B_{out/in}$, $\Psi_{out/in}$ 
and the same holds for the generators of the space translations.
\end{abstract}

\def \udt {U_D(t)}
\def \udtpm {U_{D \pm}(t)}
\def \udtmu {U_D(t)^{-1}}
\def \udts {U_D(t + s)}
\def \hdt {H_D(t)}
\def \ho  {H_0}
\def \opm {\Omega_{\pm}}
\def \omp {\Omega_+}
\def \omm {\Omega_-}
\def \li+ {\lim_{t \ra \infty}}
\def \lipm {\lim_{t \ra \pm \infty}}
\def \eht {e^{ i H t}}
\def \ut {e^{- i H t}}
\def \uas {U_{as}}

\def \udt  {U_D(t)}
\def \udts {U_D(t + s)}
\def \has  {\H^{as}} 
\def \hpm  {\H^\pm}
\def \limpm {\lim_{t \ra \pm \infty}}
\def \udtpm {U_{D \pm}(t)}
\def \udtspm {U_{D \pm}(t+s)}
\def \udtpma {U_{D \pm}^\a(t)}

\def \d {\delta}  
\def \Has {H_{as}}

\def \uast {U_{as}(t)}

\def \uint  {U_-(t)}
\def \uoutt {U_+(t)}
\def \uout  {U_+}

\def \udtst {U_D(t + s,t)}
\def \dhts  {\delta H_D(t +s)}

\def \smatrix {$S$-matrix\, }
\def \sh  {S_H}
\def \ahas {A^H_{as}}
\def \Aio {A^H_{out/in}}
\def \has {\H_{as}}

\def \ueps  {U^\eps}
\def \uepst {U^\eps(t)}
\def \uepsd {U_D^\eps}
\def \uepsdt {U_D^\eps(t)}
\def \Omepm  {\Omega^\eps_\pm}
\def \epst  {e^{- \eps |t|}}
\def \Omze {\Omega_0^\eps}
\def \Omz {\Omega_0}
\def \limp {\lim_{t \ra \infty}}
\def \veps {\tilde U^\eps}
\def \vs {\U(s)}

\def \udet {U_D^\eps(t)}
\def \we {W^{\eps}}

\def \uases {U_{as}^\eps(s)}

\def \uepsdt {U_D^\eps(t)}
\def \udt  {U_D(t)}
\def \udts {U_D(t + s)}
\def \has  {\H^{as}} 
\def \hpm  {\H^\pm}
\def \limpm {\lim_{t \ra \pm \infty}}
\def \udtpm {U_{D \pm}(t)}
\def \udtspm {U_{D \pm}(t+s)}
\def \udtpma {U_{D \pm}^\a(t)}
\def \Omepm  {\Omega^\eps_\pm}
\def \epst  {e^{- \eps |t|}}
\def \Omze {\Omega_0^\eps}
\def \Omz {\Omega_0}
\def \hedt {H_D^\eps(t)}
\def \heds {H_D^\eps(s)}
\def \cuedt {\U_D^\eps(t)}

\newcommand{\SI}{\rm{sign}}

\newpage
\section{Introduction}

For the solution of the infrared problem in quantum electrodynamics (QED) two
 strategies have been adopted: 

\noindent
A) Exploit the fact that all experiments involve limitations on
the detection of soft photons. 

\noindent In the perturbative approach, this
has led to the introduction of an infrared cutoff $\Delta E$ corresponding
to the energy resolution of the photon detectors \cite{YFS}.
A Lorentz invariant formulation of such a program has been advocated
by Buchholz \cite{Bu1} and further developed in \cite{BPS} and
\cite{BR}. A similar philosophy is at the basis of Steinmann's notion of 
particle detection \cite{St}. 


\noindent 
B) Pursue the program of the construction of an $S$-matrix 
by quantum field theory (QFT) methods. 

\noindent A crucial difficulty in this direction is the 
infraparticle spectrum of the charged particles, which seems
to preclude the existence of asymptotic Lehmann-Symanzik-Zimmermann 
(LSZ), or Haag-Ruelle limits of the charged 
fields. In fact, as a consequence of Gauss law,
the energy-momentum spectrum of the charged states cannot have
a sharp mass \cite{Bu2}, so that Dybalski's extension
of the Haag-Ruelle theory \cite{Dy} does not apply; 
in \cite{Fr}\,\cite{FMS},
such a mass spectrum has been shown to arise from 
the infinite photon content of the charged states.

Chung has proposed \cite{Ch} that the infrared divergencies
disappear in perturbative QED if a proper description of the
asymptotic states is adopted, based on non-Fock coherent factors
for the electromagnetic field, indexed by the asymptotic 
momenta of the charged particles. 
The effectiveness of the Chung ansatz for the cancellation 
of the infared divergencies has been controlled by Kibble \cite{Ki};
however, Chung ansatz raises many problems in relation with 
general structures in QFT:

\noindent
i) the space-time translation covariance of Chung's asymptotic 
charged states (depending on non translation invariant
coherent factors)

\noindent
ii) the possibility of obtaining the Chung charged states
from appropriate asymptotic charged fields

\noindent
iii) the space-time covariance and canonical structure of such fields

\noindent
iv) the existence of modified LSZ formulas for them.


A simple source of information has been provided by 
infrared models in which a semi-classical treatment is made possible by the
use of external currents or of a dipole approximation 
\cite{BN}\,\cite{Bl}\,\cite{Ki}. 
Historically, such models played a crucial role for supporting
the Chung picture; 
however, the absence of a dynamical description of charged particles
and the dipole approximation in the photon interaction 
prevent space-time translation invariance and do not allow for 
the discussion of i)-iv).

The action of the space-time translations on asymptotic
states of Chung type has been discussed in \cite{Fr}\, 
\cite{FMS} in terms of a splitting of the total Hamiltonian
$H$ and momentum $P$ as a sum, $H = H_{0 \, ph} + H_{charge} $, 
of a free photon term and a ``particle'' contribution, with
definite mass. In this picture, the infraparticle spectrum is
explained by the infinite photon content of the charged states.

A non-perturbative control of the validity of Chung ansatz
has been obtained in non-relativistic QED \cite{CFP}. The
main difficulty in non-relativistic models is given by the electromagnetic 
corrections to the energy-momentum relation $E(p)$ of the
asymptotic charged particles, a problem which plays a minor role
in QED, where, by the Lorentz invariance of the energy-momentum spectrum
(which persists for charged states \cite{BB} in spite of the non 
implementability of the Lorentz symmetry),
only a mass enormalization is admitted. 
An LSZ modified formula for one charged particle states
has been derived in \cite{CFP}; the discussion of asymptotic  
particle observables has been restricted to the velocity 
and Coulomb effects have not been discussed.


\vspace{1mm}
A concrete approach to the complete construction of an
$S$-matrix in QED in terms of M{\o}ller operators has been discussed by
Kulish and Faddeev \cite{KF} and by Rohrlich \cite{JR},
who realized the crucial role of Dollard strategy 
of using a modified large time dynamics \cite{Do1}\,\cite{Do2},
rather than the standard free dynamics.

Dollard strategy has been exploited and rigorously controlled only
for non-relativistic Coulomb scattering and its extension to QED faces
substantial problems, due to the infinite photon emission, 
the related persistent effects and the lack of  covariance under space-time 
translations of the photon interaction in Dollard's dynamics
(beyond the choice of the initial time in Dollard treatment of Coulomb 
interaction).

On the basis of the Kulish-Faddeev identification of a Dollard reference
dynamics in QED, Rohrlich has proposed \cite{JR} to use it with the same role 
as the free dynamics in the standard interaction picture. 
In this way, he introduces (non free) fields and states 
associated with a ``modified interaction picture'' and describes the  
scattering in terms of transitions between such states; he also argues that, 
for large times, such fields coincide with the asymptotic fields proposed by 
Zwanziger \cite{Z}. 
The time dependence of such asymptotic fields, reflecting 
asymptotic effects of photon emission and Coulomb distorsions,
does not satisfy the group property and space-time covariance 
is problematic.




\vspace{1mm}
As before, soluble infrared models do not provide instructive information 
on such issues: in particular, the Pauli-Fierz-Blanchard 
model \cite{Bl} leads to a Dollard dynamics which is a group and
substantially includes the full photon interaction.

\vspace{1mm}
The aim of this work is to shed light on 
the above problems by discussing a model describing Coulomb 
scattering with photon emission by $N$ heavy charged particles, treated as 
classical particles, with no recoil induced by the photon interaction.
Even if the particle dynamics is not affected by the photon interaction,
the model is not explicitely solvable, the photon dynamics depends on the
non-trivial particle trajectories and detailed estimates are needed
for the control of infrared effects arising from Coulomb 
asymptotic distorsions. 

Such distorsions are of the same form ($\sim \log t$) as those 
induced by photon recoil, which should not therefore lead to
substantial changes for the control of the asymptotic dynamics.
Moreover, the results of \cite{CFP} confirm that photon recoil
does not change the (coherent) characterization of asymptotic
photons in terms of the asymptotic charged particle velocity. 

As a consequence of the no-recoil approximation, there is no scattering
in the one charged particle sector; however, in the many particle sectors
the $S-$matrix is non-trivial and for its construction the Coulomb asymptotic
distortions play a substantial role. The overall picture qualifies as
a realistic description of scattering of heavy charged particles,
with strong effects for particles of high charge.
 
The fact that the interaction is translation invariant and the 
electromagnetic current is not pre-assigned is a distinctive feature of the
model with respect to the external current and Pauli-Fierz-Blanchard models.
In fact, the model allows for the discussion of the asymptotic limit 
of charged particle variables and charged fields, with non-trivial
electromagnetic effects even in the one charged particle sector and
a complete control of Coulomb effects. Asymptotic field
algebras are constructed, with a full control of the problems 
raised in i)-iv).

The model is defined by the Hamiltonian
$$ H = h_0 + h_I + H_0 + H_{I, \, r} \, ,$$
where $h_0$ is the (non-relativistic or relativistic)
free Hamiltonian of $N$ classical particles in the Koopman formulation,
$h_I$ the Coulomb interaction,
$ H_0 $ the free photon Hamiltonian,     
$ H_{I , \, r} $ the (renormalized) particle-photon interaction.
$H$ is invariant under space translations, with generator denoted by
$\P$.

The first aim is to check the effectiveness of the Dollard-Kulish-
Faddeev-Rohrlich strategy for the mathematical control of the asymptotic 
limit. In accord with the general 
approach and analysis of \cite{MS}, we obtain, for the 
$N$ charged particle channel (see Theorem 5.4):

\noindent
1) {\em Existence of the M{\o}ller operators}. The M{\o}ller operators 
$\Omega_{\pm}$  are obtained by introducing 
an adiabatic switching, $e^{- \eps |t|}$, of the 
electromagnetic coupling, and a Dollard 
reference dynamics $U_D^\eps(t)$  
\be{ \Omega_\pm = \lim_{\eps \to 0 , \, t \to \pm \infty} 
U^{\eps \, *}(t) \, U_D^\eps(t) \, .}
\ee


\noindent
2) {\em The group of asymptotic space-time translations}. 
There is a unique family of unitary operators 
$U_{as}(a,t)$, 
$ a \in \dR^3$, $ t \in \dR $, 
satisfying the interpolation formulas, eq\,(1.1), and
therefore the group property and strong continuity, given by
\be {
U_{as}(a,s) \eqq
U_+(a, s) = U_-(a,s) =  \lim_{\eps \to 0 \, , \,t \to \pm\infty} 
U_D^{\eps \, *}(t) \, U_0(a,s) \, U_D^{\eps}(t) \, , }
\ee
with $U_0(a, t)$ the free space-time translations. 
The corresponding generators are 
$$ H_{as} = h_0 + \alpha_{as} (H_0 (a^*, a)) \ , \, \, \,
\P_{as} = - i \sum_i^N \partial/ \partial q_i + 
  \alpha_{as} (P_{ph}(a^*,a))\, ,
$$
with $P_{ph}$ the photon momentum and  $\alpha_{as} $ 
the standard non-Fock coherent shift, 
$a^* \to a^* + \sum_i J(p_i)$, associated to the momenta $p_i$ of the charged 
particles, eqs.\,(4.17),\,(5.18). 
$U_{as}$ is determined by $U_D$, but the two notions are
basically different, contrary to the discussion of the
asymptotic fields adopted in Refs.\,\cite{Z}\, \cite{JR}.

\noindent
3) The {\em $S$-matrix}, $ S = \Omega_+^* \, \Omega_{-}$ is {\em invariant under 
the asymptotic space-time translations} $U_{as}(a,t)$. 

\noindent 
4) {\em The infrared divergences} due to the Coulomb interaction and to the 
soft photon emission {\em factorize}, eq.\,(5.27).

\vspace{1mm}
The second aim of this work is to shed light on 
the asymptotic limits of the Heisenberg fields, expecially
on the still open problem of the existence of LSZ limits of charged 
fields and of their space-time covariance properties.

This can be done in the (second quantized) field theory 
version of the model, with the introduction of charged 
fields $\Psi^*(f, t)$, $f$  
a test function of the Koopman variables. 
In spite of the no-recoil approximation, 
such charged fields are space-time covariant, with non-trivial 
and not pre-assigned dynamics; their asymptotic limit
exists as a strong limit, providing 
information which is not available in soluble infrared 
models \cite{BN}\,\cite{Bl}\,\cite{Ki}
and is only partially given by the analysis of the one charged particle 
sector in non-relativistic QED \cite{CFP}.  
 
On one side, the M{\o}ller operators automatically provide
Heisenberg asymptotic fields,  defined on the scattering 
spaces $\H_\pm$, eq.\,(5.8), by 
\be{  \Psi^*_{out/in} (f) = \Omega_\pm \, \Psi^*(f) \, \Omega^*_\pm \ , 
\ \ \ \ a_{out/in}(g) = \Omega_\pm \, a(g) \, \Omega^*_\pm
\, .}
\ee
The fields $\Psi^*_{out/in}$ and $a_{out/in}$ obey (equal times) canonical 
commutation relations, but their (Heisenberg) time evolution is 
not free, being explicitly given by 
$H = H_{as} (\Psi_{out/in} , \, a_{out/in}) $. 
This shows the effectiveness of the Kulish-Faddeev-Rohrlich-Zwanziger
strategy also for the construction of asymptotic fields; their dynamics
is not free, but no Coulomb distorsion appears (in contrast with
the properties of the asymptotic fields proposed by 
Zwanziger and Rohrlich).

For the existence of the asymptotic limit of charged fields, the
Dollard corrections are essential; they give rise to 
{\em modified LSZ (Haag-Ruelle) formulas}.
In fact, for the charged Heisenberg fields in the variables 
$P = - i \partial/\partial q$ , $p$ (see eq.\,(6.1)), 
one has 
$$ \Psi^*_{out/in} (f) = \lim_{\eps \to 0} \lim_{t \to \pm \infty}
  \int dP \, dp \, f_{-t}(P,p) \, \Psi^{\eps \, *}_t(P,p) \, 
 e^{i \rho^\eps_t \, (\chi^\eps_t(P,\,p))} $$
\be { \times  \exp{-i  \int_0^t ds \,   
 A^{\eps}_t 
( \stackrel{\leftrightarrow}{\partial_t} 
D_{t-s} * j^\eps(v(p); s))} 
\, , }
\ee
where 
$f_t$  denotes the free time evolution of $f$, 
$A$ is the electromagnetic potential, $D_t(x)$ the massless
commutator function, 
$\Psi^\eps_t, \rho^\eps_t,  A^{\eps}_t $ 
the Heisenberg (adiabatically switched) 
time evolution  of the corresponding variables,
${*}$ the convolution in the space variable $x$,
$ j^\eps_\mu (v(p); x, x_0) \eqq e \, v_\mu(p) \tilde \eta(x-vx_0) \, 
 e^{- \eps |x_0|} $, $v_\mu \eqq (1, v)$, a function of $p$
corresponding to the (free) asymptotic 
particle current (with an ultraviolet cutoff $\eta(k)$),  
$\rho$  is the charge density,    
$\chi$ is a Coulomb phase.

As derived in general by Buchholz \cite{Bu} on the basis of locality and of
the Huyghens principle, LSZ (Haag-Ruelle) asymptotic
limits of the electromagnetic fields exist, without any Dollard
correction, and define massless fields. 
In our case, the ordinary LSZ procedure converges in all sectors
and yields massless asymptotic fields $b_{out/in}$; 
they are related to $a_{out/in}$ by 
\be{ b_{out/in} (g) =  a_{out/in}(g) +  (\rho_{out/in} (J))(g)  \, , }
\ee
for any $g(k, \lambda) \in \S(\Rbf^3) $.
All the above asymptotic limits exist in the strong operator sense 
on the scattering space (on an invariant dense domain, for e.m. fields), 
yielding the usual Haag-Ruelle (equal times) limits. 

The fields $\Psi_{out/in}$ and $b_{out/in}$ define a {\em semi-direct product 
of canonical algebras}, with the non-standard commutation relations
($b^\# = b, \, b^*$)
\be {
[ b^\#_{out/in}(k, \lambda), \,  \Psi^*_{out/in} (P,p) ] = 
   J (k, \lambda, p) \, \Psi^*_{out/in} (P,p)  \, }
\ee
and 
$  H = h_0 (\Psi_{out/in}) + H_0 (b_{out/in}) $
on the scattering space; 
while $b_{out/in}$ are free fields, $ \Psi_{out/in} $ are not.

The LSZ (Haag-Ruelle) formula for the charged fields can also be written
with the e.m. factor replaced by \emph{the exponential of a string-like
integral of the asymptotic photon field}, eq.\,(6.24).

Equation (1.4) corresponds to the following LSZ
formula for QED, which automatically arises from the Dollard 
dynamics of the charged fields introduced by Kulish-Faddeev and
Rohrlich (KFR), through the same steps as in Sects. 6.1, 6.2a:
$$ \psi^*_{out/in} (f) = \lim_{\eps \to 0}  \lim_{t \to \pm \infty}
  \int  d^3p \,\, f_{-t}(p) \, \psi^{\eps \, *}_t(p) \,\,
e^{i \frac{e^2}{4 \pi} \, {\SI} \,t \,\ln |t| \, \int d^3q \, \frac {\rho(q)} {v(p,q) } \, } 
$$
\be { \times  \exp {-i  \int_0^t ds \,   
 A^{\eps}_t 
( \stackrel{\leftrightarrow}{\partial_t} 
D_{t-s} * j^\eps(v(p); s))} 
\, , }
\ee
with $v(p,q)$ the Lorentz invariant relative velocity and
$j^\eps_\mu (v(p); s, \x) \equiv $ \ $e \, v_\mu(p) \, \delta (\x - {\bf v} s)$,
see eqs.\,(S4-10,11), (S4-21,22) in \cite{JR}.

The effectivness of the KFR strategy in QED has been controlled by 
Zwanziger \cite{Z1}, with the cancellation of the infrared 
divergencies in the perturbative expansion of the reduction formulas.
Our model gives non-perturbative support to that strategy; it
also shows that the same strategy leads to asymptotic fields as 
(modified) LSZ (HR) strong limits, with a complete control of the ensuing
structure and covariance properties of the asymptotic field 
algebras. The space-time covariance properties
of the asymptotic fields, eq.\,(6.19), 
and their commutation relations, eq.\,(1.6), directly 
follow from the last (e.m. field) term in eq.\,(1.4);
eq.\,(1.7) gives rise to the same relations, with 
$ \Psi^*_{out/in} (P,p)$ replaced by $ \psi^*_{out/in} (p)$. 

In both eqs.\,(1.4),\,(1.7) the Dollard modifications
are parametrized by $v(p)$, $p$ the momentum variable of the 
interacting Heisenberg field, smeared with $f_{-t}$;
convergence of the LSZ formula implies that $v(p)$ can be 
identified with the asymptotic ``particle'' velocity,
independently of recoil assumptions.  

The use of an adiabatic cutoff, while technically important,
does not seem to be essential for the results since, in the model, 
one may avoid it by adopting a modified (Dollard) 
correction to the LSZ formula for the charged fields \cite{MSpr}.

\vspace{1mm}
The lesson for the infrared problem in QED is manifold, 
briefly:

\noindent 
i) the Kulish-Faddeev-Rohrlich approach to the infrared problem in QED
\cite{KF}\,\cite{JR}, based on Dollard's strategy, allows for a 
systematic control of the asymptotic limit;

\noindent
ii) in particular, it allows for the 
reexamination of the open problem of 
the asymptotic condition for charged fields;
modified LSZ (HR) formulas can be written, yielding asymptotic
limits of charged fields;

\noindent
iii) the asymptotic field algebra, 
generated by a free photon field (given by Buchholz asymptotic limit)
and by the asymptotic charged fields,
has the structure of a semidirect product, reproducing Chung's 
ansatz; 

\noindent
iv) the Hamiltonian is the sum of the  
free Hamiltonians of the asymptotic fields, eq.\,(6.17); 
the time evolution of the asymptotic charged 
fields is not free only for the presence of infrared photons, 
with no residue of the Coulomb interaction.

\goodbreak



\section{The model}

The model describes $N$ classical charged particles of 
charge $e_i$ and mass $m_i$, $i =1,...N$, with mutual interaction given by  
a Coulomb potential $\V$, regularized 
at the origin, and interacting with  (transverse) photons. 

The classical particles configurations are described by wave functions 
on the phase space $\Gamma $, $\psi(\q_1, ...,\q_N; \p_1, ...,\p_N) 
\in L^2(d^{3N} q, d^{3N} p)$, 
$|\psi|^2$ representing the density in phase space governed by the
Liouville time evolution; the time evolution of $\psi$ 
is given by the Koopman Hamiltonian
\be{h = -i \sum_i \left ( v_i \, \frac{\partial}{\partial q_i} - 
\frac{\partial \V}{\partial q_i} \frac{\partial}{ \partial p_i} \right ) 
\eqq h_0 + h_I \, ,}\ee
\be {
   \V(q) = \sum_{j \neq i}  
\frac{ e_i e_j}{8 \pi (|q_i - q_j|^2 + a^2)^{1/2}} \, .}\ee

\noindent
We consider both the non-relativistic and the relativistic case, 
given by
\be{ v_i(p) \eqq p_i/m \, , \ \ \mbox{respectively} \ \
  v_i(p) \eqq \frac{p_i}{(p_i^2 + m^2)^{1/2} } \, . }
\ee
For simplicity, we have omitted the  vector notation for the operators 
$\q_i$, $\p_i$, $\mathbf{\partial/\partial q}_i$, 
$\mathbf{\partial/\partial \p}_i$ 
and for their scalar products; this will also 
be done in the following.
 $q_t(q, p), \,p_t(q, p)$ 
will denote the solutions of the classical equations, with initial data $q,p$.


The Hilbert space $\H = L^2(\Gamma) \times \H_F$ 
can be identified with the space of $L^2$ functions 
$\psi(q, p)$, $ q = (q_1, ...,q_N)$, $p =(p_1, ...,p_N)$, 
taking values in the Fock Hilbert space $\H_F$, where  
the transverse photons are described by the standard canonical destruction and 
creation operators $a(k,\l)$, $a^*(k,\l)$ ($\l = \pm 1$ denoting the helicity), 
 with commutation relations 
$ [\,a(k,\l), \,a^*(k',\l')\,] = \d(k-k')\,\d_{\l\,\l'}, \,\,\,\mbox{etc.} $

\noindent
The total Hamiltonian $H$ is 
(always omitting the vector notation and 
denoting by $\epsilon (k,\l)$ the polarization vectors)
\be{H = h + H_0 + H_{I,\, r}, \ \ \
H_0 = \sum_{\l = \pm 1}\int d^3 k \, |k|\,a^*(k,\l)\, a(k,\l), }\ee
\be{  H_{I,\, r} = H_I - \Delta E(p),  \ \  H_I =   
a(f(q, p)) + a(f(q, p))^* \eqq  H_I(a, a^*, q, p), }\ee 
\be {f(k,\lambda;q,p) = 
\frac{1}{(2 \pi)^{3/2}}  \sum_i 
\frac{1}{\sqrt{2 |k|}}  e_i 
\eta(k)\, e^{ i k q_i}\, \epsilon (k,\l)\,v_i(p) }\ , \ee
$$ a(f(q, p)) \eqq \sum_{\l = \pm 1} \int d^3 k  \, a(k,\l)
  f(k, \l ; q,p) \, , $$
where we have introduced a (real rotationally invariant) 
ultraviolet cutoff $\eta(k)$, $\eta(0) = 1$, 
with Fourier transform  
 $\tilde{\eta} \in    \D(\Rbf^3)$;    
 $\Delta E(p)$ is a $C^1$
function of the particle momenta, to be determined below (see Section 5), 
playing  the role of a mass counter-term, 
subtracting persistent effects of $H_I$. 

The Hamiltonian $H$ is invariant under space 
translations, $T(a)$, $a \in \Rbf^3$ and the corresponding generator  
\be {
 \P = - i \sum_i \partial/ \partial q_i  + P_{ph}, 
\,\,\,\,\,P_{ph} = \sum_{\l = \pm 1}\int d^3 k \, k\,a^*(k,\l)\, a(k,\l) \, ,}
\ee
is conserved. $P_c = \sum_i p_i $ is also conserved, but $ P_c + P_{ph} 
\neq \P$ is not, corresponding  to the absence of particle 
recoil in the photon emission.

The particle total energy
$$ E(q, p) = K(p) + \V(q) \, ,$$
$K(p)$ the kinetic energy,
commutes with $H$, and is therefore a constant of motion; however, 
$E + H_0 + H_{I, \, r}$ is not.

For the treatment of photon emission, it is important  to limit the 
particle velocities to be smaller than the velocity of light 
(which is 1 in our units). To this purpose, in the non-relativistic case,
we shall restrict 
our discussion to a suitable subspace $\H_{nr} \subset \H$.

Assuming that
$$ \kappa_0 \eqq \frac{ N^2 e_{max}^2}{4\pi a\, m_{min}} < \ume \ ,$$
it is enough to take 
\be{\H_{nr} = P_{nr} \H \eqq L^2(\Gamma_{nr}) \times \H_F \, ,}
\ee
with $P_{nr}$ the projector on 
$K \leq K_{max} \eqq \ume \kappa \, m_{min}$, with $\kappa < 1 - 2 \kappa_0$. 
In fact, the conservation of $E$ implies 
\be
{ v_{i, \,t}^2 \leq (\kappa + 2 \kappa_0) < 1, \ \ 
\forall t, \ \forall i\ . }
\ee

Independently of the infrared problem, the convergence of the
M{\o}l\-ler operators requires some ``time smearing'' and to
this purpose we adopt an adiabatic regularization given by an 
adiabatic switching 
$e^{-\eps |t|}$. It can be combined with
the Dollard strategy, as discussed in \cite{MS}. It is enough to use
it only for the particle-photon interaction; its introduction also for
the Coulomb interaction will only be convenient for displaying a complete 
factorization of the infrared divergences.

As needed in all QFT models with persistent effects, mass counterterms
will be introduced, both in the Hamiltonian and in the Dollard correction 
to the free dynamics.

\vspace{1mm} 

In the $N$ charged particle sector, 
we always consider the $N$ particle channel, corresponding
to asymptotic configurations excluding charged particle clusters.
All the following results hold for all values of the charges
and of the masses; for simplicity, in the following, we shall omit 
the particle indices for charges and masses, 
putting $e_i = e$, $m_i = m$. 

\vspace{1mm} 
We start with the particle dynamics and scattering,
providing the M{\o}ller operators in the Koopman formulation and
the estimates on the particle trajectories which are 
needed for the control of photon emission (Section 3); 
for both purposes, the main point is the control of the 
Coulomb effects (the analysis becoming much simpler, 
but still instructive, for short-range potentials).  

Then we introduce the (renormalized) electromagnetic interaction and 
a Dollard reference dynamics for the full time evolution, both with an 
adiabatic regularization, and prove the existence of the 
corresponding M{\o}ller operators (Section 4, Proposition 4.1). 

In Section 5 we discuss the removal of the adiabatic switching, 
after the specification of the mass 
counterterm, obtaining the existence of the  
M{\o}ller operators (Section 5.1), the existence and characterization 
of the asymptotic dynamics (Section 5.2), the interpolation formulas 
and the explicit factorization of the infrared divergences 
in the $S$-matrix (Theorem 5.4). 

In Section 6, we discuss the second quantized version of the model and prove 
the existence of  LSZ asymptotic limits of the Heisenberg 
fields (Section 6.2a,b);  the asymptotic algebra  and its covariance under 
space time translations are analyzed  in Section 6.2c. 
Finally, in Section 6.2d we derive an asymptotic form of the corrections  
to the standard LSZ formulas for the charged fields.   


\newpage

\section{Particle dynamics and scattering}

For the definition of the Dollard reference dynamics for the particles, 
it is convenient to introduce the following operators:
\be{ Q_i \eqq  i \partial/\partial p_i, \,\,\,\,\,\,\,\,\,P_i 
\eqq - i \partial/\partial q_i, }\ee
which satisfy
\be{ [\,q_i, \,P_i\,] = i \delta_{i\,j}, \,\,\,\,\,\,\,[\,p_i,\, Q_i\,] 
= - i \delta_{i\,j}.}\ee
Then, in a notation covering both the non-relativistic 
and the relativistic case, 
\be{ h_0 = \sum_i v_i P_i \, , \,\,\,\,\,\,\,\,\,h_I = 
\sum_{i\,j; j \neq i} w_{i\,j}(q;a) Q_i \eqq h_I(q, Q; a)\, ,}\ee
$$ 
w_{i\,j}(q;a) = 
\frac{- e_i e_j\,(q_i - q_j)}{4 \pi (|q_i - q_j|^2 + a^2)^{3/2}} \, . $$

The free Heisenberg evolution is
$$ q_i(t) = q_i + v_i t, \,\,\,\,\,\,p_i(t) = p_i, \,\,\,\,\,\,Q_i(t) 
= Q_i + V_i t, \,\,\,\,\,\,P_i(t) = P_i,$$
with ($\alpha, \beta = 1,2,3$ the vector components)
\be{ V_i = (V_i^\alpha(p, P), \, \alpha =1, 2,3), \,\,\,\,\, V_i^\alpha(p, P) 
\eqq \sum_{\beta=1}^3 
\frac{\partial v_i^\beta(p)}{\partial p_i^\alpha} P_i^\beta \, , }\ee
 reducing to $ V_i = P_i/m $ in the non-relativistic case.

With the same motivations as in Dollard treatment of Coulomb 
scattering, \cite{Do1}\,\cite{Do2}\,\cite{MS}, 
a reference large time dynamics for 
particle scattering may be identified by putting $q = vt$,
$Q = Vt$ in $h_I$, where for simplicity we take $a =0$:
\be{h_D(t) \eqq  h_0 + h_I(v t, V t; 0) \, .  }
\ee
No adiabatic switching is necessary for  the particle scattering 
and for simplicity
it will not be introduced until the end of Section 5, 
where it will be used in order to display     
the explicit dependence on the ``infrared cutoff'' $\eps$ in the $S$-matrix.

\begin{Proposition} 1) The Hamiltonian $h$ is essentially self-adjoint  
on the domain $C_0^1 \subset L^2(\Gamma)$ of 
differentiable functions $\psi(q, p)$ 
of compact support and its exponential $u(t) = e^{- i h t}$ leaves 
$C_0^1$ invariant;

\noindent 
2) the equation
\be
{ i d u_D(t) /dt = h_D(t)\, u_D(t)\, , \ \ \ u_D(\pm 1) = u_0 (\pm 1) \, , }
\ee
with $u_0(t) \eqq e^{ - i h_0 t}$ and $h_D(t) $ given by eq.\,(3.5), 
has a unique solution for $|t| \geq 1$
leaving  $C_0^1$ invariant, given by 
\be{ u_D(t) = u_0(t) \exp \left(  i \frac{e^2}{4 \pi}\, {\SI} \,t \,\ln |t| 
\sum_{i<j}  \frac{v_i - v_j}{|v_i - v_j|^3} 
( V_i - V_j) \right) \, ,}
\ee
satisfying 
\be{ [\,u_0(s) ,\,u_D(t)\,] = 0,\,\,\,\,\,\,[\,p, \,u_D(t)\,] = 0, 
\,\,\,\,\,\,[\,P, \,u_D(t)\,] = 0;}\ee

\noindent 3) the following strong limits exist
\be{ \mbox{strong}-\limpm u(t)^*\,u_D(t) = \omega_\pm;}\ee
moreover
$$ \mbox{strong}-\limpm u_D(t)^* \,u_D(t+s) =  u_0(s),$$ 
\be{ u(t)\,\omega_\pm = \omega_\pm u_0(t)\ ;}\ee

\def \coud   {C_0^1(\delta, K)}

\noindent 4) let $C_0^1(\delta, K)$ denote the set of $\psi(q, p)$ 
with support in a compact set $K$ and such that $\psi(q, p) = 0$ 
if $|v_i - v_j| < \delta$, for some $i \neq j$,   
then, $\forall \psi \in D_\pm  \eqq \omega_\pm
 D^1_0$,  $D^1_0 \eqq 
\cup_{\delta\,K} \, C_0^1(\delta, K)$ dense in $L^2(\Gamma)$,
\be{||(p_t - p_\pm)\,\psi||_{L^2} = O(|t|^{-1} \ln|t|),}\ee
where  $p_\pm$ is defined, on $D_\pm$, by 
\be{p_\pm\, \om_\pm = \om_\pm \,p \, ;} \ee 

%


\noindent 5)  
let $\Gamma_\pm $ be the complements of the sets 
$$\{(q,p): \omega(t)\psi(q,p) \to 0, \, t \to \pm \infty, \, 
\forall \psi \in D_0^1 \}  \, , \ \ 
\omega(t) \eqq u^*(t) u_D(t) \,  ;$$ 
then, $\om_\pm L^2(\Gamma) = L^2(\Gamma_\pm)$ and
$ (\omega_\pm \psi) (q,p) = \psi (\gamma_\pm (q,p)) 
\, , \  (q,p)\in \Gamma_\pm \, , $
with $\gamma_\pm : \Gamma_\pm \mapsto \Gamma = L^2(d^{3N} q, d^{3N} p)$ 
 measure preserving transformations. $p_\pm$ are multiplication
operators $p_\pm(q,p)$ on $L^2(\Gamma_\pm)$, essentially self-adjoint  
on $D_\pm$ and,
almost everywhere in $\Gamma_\pm$ (with $v_\pm \eqq v(p_\pm$),
\be { p_t(q,p) = p_\pm + \delta p_\pm \, t^{-1} + O(t^{-2}\ln t)  \ ,
\ \  \delta p_{\pm \, i} \eqq - \sum_ {j \neq i} w_{ij} (v_\pm, 0) ; }\ee

\noindent 6)  
$\om_\pm  K = (K + \V) \om_\pm$ and therefore
$s \eqq  \om_+^* \om_{-} $ 
commutes with the particle kinetic energy
$K$; thus, in the non-relativistic case, it leaves  
$\H_{nr}$ invariant.

\end{Proposition}
\def \Cou  {C_0^1}

\noindent
\Pf\, 1). The solutions of the classical equations  $q_t(q, p)$, $p_t(q, p)$ 
are $C^1$ functions of $q, p$ and $t$, actually $p_i$ are 
uniformly bounded in $t$, as a consequence of the energy conservation 
($E = K + \V(q)$, $\V$ bounded below); hence they define a 
one-parameter unitary group 
\be {  u(t) \psi(q, p) \eqq \psi(q_{-t}(q,p), p_{-t}(q,p))\, , }
\ee
with locally finite propagation speed, leaving 
$\Cou$ invariant. By Stone theorem, its generator is e.s.a on  
an invariant domain.

\noindent 2). Since the argument of the exponential in eq.\,(3.7) 
has  a dense invariant domain 
of analytic vectors, 
given, e.g., by functions $\psi(q, \,p)$ analytic in $q$ and of 
compact support in $p$, with $|v_i - v_j| > \delta$, $\forall i \neq j$,  
the right hand side of eq.\,(3.7) 
is well defined and 
\be{ u_D(t) \psi(q, p) \eqq \psi(q_{-t}^D, p_{-t}^D)\, , }
\ee
$$ (q_t^{D\alpha})_i \eqq q_i^\alpha + v_i^\alpha t - \frac{e^2 }{4 \pi} 
{\SI}\,t \ln |t| \sum_{j \neq i} \frac{(v_i - v_j)^\beta}
{|v_i - v_j|^3}\frac{\partial v_i^\alpha}{\partial p_i^\beta}   
\ , \ \ \ (p_t^D)_i \eqq p_i \, ,$$
leaves $\Cou$ invariant and  satisfies eq.\,(3.6).

\noindent By hermiticity of $h_D(t)$ on $\Cou$, for any two solutions 
$u_D^1$, $u_D^2$, one has 
$$ (d/dt)(u_D^1(t) \psi,\,u_D^2(t) \psi) = 0$$
and uniqueness follows. 

\def \coud   {C_0^1(\delta, K)}

\noindent 3). For the existence of $\om_\pm$, we note that 
$\forall \psi \in \coud$,  using 
$$ u_D(t)^*\, Q \,u_D(t)  = Q + V t \eqq Q_t^D $$
and eqs.\,(3.8), one has 
$$ i (d/dt) (u^*(t) \,u_D(t)) \psi = u^*(t) 
\,[ - h_I(q, Q) + h_I(v t, \,V t; 0) ] u_D(t)\,\psi = $$
$$= u^*(t)\, u_D(t) [ - h_I(q_t^D, Q_t^D) + h_I(v t, V t ; 0)]\,\psi = $$
$$ = u^*(t)\,u_D(t) 
\sum_{i \neq j} [ - w_{ij}(q_t^D)\, Q_i + \left ( - w_{ij}(q_t^D) 
+ w_{ij}(v t ; 0)\,\right ) V_i t\,] \psi\, .   $$
Now, using $ || \partial v_i^\beta/\partial p_i^\alpha|| \leq m^{-1} $,
\be {  |q_{i, t}^D - q_{j, t}^D|  \geq \delta |t| - 
2 (e^2 /4 \pi m)(N-1) \delta^{-2}  \,\ln |t| -  
\,\mbox{sup}_{K, i, j} |q_i - q_j| \, , }
\ee
so that  the norm of first term is bounded 
by $O(|t|^{- 2})$. The same holds for
the Sup norm, since $u^*(t)\,u_D(t) $ amounts to a change of variables.

\noindent The difference in round brackets consists of a term which 
can be estimated by $O(|t|^{- 2} \, \ln |t|)$
and a term of the form 
$ (v_i - v_j) t \left (a_t^{-3} -  b_t^{-3} \right)$,
with, for large $|t|$,  
$$ a_t = O(|t|), \,\,\,\,\, b_t = O(|t|), \,\,\,\,\,
|a_t - b_t| = O(\ln |t|).$$
Therefore, the term is bounded by $O(|t|^{-2} \ln|t|)$. 

\noindent In conclusion, one has  
\be   
 { || (d/dt) u^*(t) \,u_D(t) \psi|| \leq  O(|t|^{-2}\,\ln|t|) \, .}
\ee
and the same holds for the Sup norm.
The first of eqs.\,(3.10)  follows immediately from the 
explicit form of $u_D(t)$; then the second follows as in 
Propositions 2.1, 2.2 of \cite{MS}.

\noindent 4).  
The above estimates imply,  
$\forall \psi \in C_0^1(\delta, K)$, pointwise convergence of
$ (\omega(t) \psi) (q,p)$, for $|t| \ra \infty$, and  
$$||(\om_\pm - \om(t))\,\psi || = O(|t|^{-1}\,\ln |t|).$$
Therefore, on $C_0^1(\delta, K)$, where the multiplication operators 
$p_t = p_t(q,p) $ are bounded uniformly in $t$, 
using $p_t = \om(t)\, p\, \om^*(t)$, one has
$$p_t\,\om_\pm = p_t\,(\om(t) + O(|t|^{-1}\,\ln|t|)) = 
\om(t)\,p + O(|t|^{-1}\,\ln|t|) =$$ $$= \om_\pm \,p + O(|t|^{-1}\,\ln|t|)  = 
p_\pm\,\om_\pm + O(|t|^{-1}\,\ln|t|) \, . $$  
\def \cuet {\U^\eps(t)}
\def \cuedt {\U_D^\eps(t)}

\noindent 5). 
Lemma A.1 applies with $\omega_t \to \omega(t)$ and $D \to
\cup_n C_0^1(1/n, K_n)$, for any sequence $K_n$ covering $\Gamma$
and $\om_\pm L^2(\Gamma) = L^2(\Gamma_\pm)$ follows.
In the notation of Lemma A.1, $p_t(q,p) = p (\gamma_t(q,p))$, 
which converges to $p(\gamma_\pm(q,p)) = p_\pm(q,p)$ for 
almost all $ (q,p) \in \Gamma_\pm$; then,
by eq.\,(3.11), $\forall \psi \in D^1_0 $, 
$$(p_\pm \omega_\pm \psi) (q,p) = \lim_{t \ra \pm \infty} 
(\omega(t) p \psi) (q,p) = 
\lim_{t \ra \pm \infty} p(\gamma_t(q,p)) \psi (\gamma_t(q,p)) $$
$$ = p_\pm(q,p) (\omega_\pm  \psi)(q,p). $$
Therefore, $p_\pm$ coincide with the multiplication operators
$p_\pm(q,p)$ on $D_\pm$, where $p_\pm$ are e.s.a. by eq.\,(3.12).
Convergence of $p_t(q,p)$ to $p_\pm(q,p)$ on $\Gamma_\pm$ and
\be{q_t(q, p) - q = \int_0^t d s\, v(p_s) }\ee
imply, for $(q,p) \in \Gamma_\pm$, $\forall \eps >0$, for large $|s|$, 
$$  |q_{i, s}(q, p) - q_{j, s}(q, p)| \geq (1-\eps) 
|v_{i \, \pm}(q, p) - v_{j \, \pm}(q, p)| \, |s|\ ,  \ \ \forall i \neq j  
\,  $$
with $v_\pm(q,p) \eqq v(p_\pm)$. 
By eq.\,(3.12), $|v_{i \, \pm} - v_{j \, \pm}| \neq 0$ a.e. in 
$\Gamma_\pm$; hence, a.e. in $\Gamma_\pm$,
$$
  t^2 \, d p_{i\,t}/dt = - t^2 \sum_{j \neq i} w_{i\,j}(q_t, a) =
  -  \sum_{j \neq i} w_{i\,j}(q_t/t, a/t) $$
\be {\to - \sum_{j \neq i} w_{i\,j}(v_\pm, 0) 
\eqq  \delta p_{i, \pm}. \ ,}\ee
This implies $v(p_s) = v_\pm + O(1/s)$ and,
by eq.\,(3.18),
$$ q_t / t - v_\pm = O(\ln t /t) \ , $$
which gives 
$$ t^2 \, d \, p_{i\,t}/dt = - \delta p_{i\, \pm} (q,p) + O(\ln t /t) \, , $$
a.e. in $\Gamma_\pm$;
eq.\,(3.13) then follows from
$$ p_{i\, \pm} - p_{i\,t}  = \int_t^\infty d s\,\, d p_{i\, s}/d s .$$ 

\noindent 6).
$ u_D^* (t) \,\V(q)\, u_D(t) 
= \V(q^D_t) $ is bounded uniformly in $t$ and 
converges strongly to zero by eq.(3.16). Therefore, since 
$[ K + \V, \,u(t)^*\,] = 0$ and by eq.\,(3.8),
$[ K, \, u_D(t) ] = 0$,   
$$ \om_\pm  K = (K + \V) \om_\pm - 
\lim_{t \to \pm \infty} u(t)^* u_D (t)  \V(q^D_t)  =  
(K + \V) \om_\pm \, .  \ \endproof $$ \goodbreak

In the case of repulsive Coulomb potential, the unitarity of $\omega_\pm$
easily follows for $N = 2$; a proof is given in Appendix B. \goodbreak





\section{Dynamics and Scattering with an Adiabatic Regularization}
 
We shall construct a regularized dynamics $U^\eps(t)$ and a 
regularized Dollard reference dynamics $U_D^\eps(t)$, corresponding to the  
the  substitution: $q_i \ra v_i t,$  in $H_{I. \, r}$; we choose an
$\eps$ regularization corresponding to the replacement $e \to e^{-\eps |t|} e$:
\be {H_{I,\, r} \to H_{I, \,r}^\eps (t)  \eqq  e^{-\eps |t|} H_I - e^{- 2 \eps |t|} 
\Delta E(p) \ ,}
\ee
\be
{H_{I D}^\eps(t) \eqq  e^{- \eps |t|}\, H_I(a, a^*, v t, p) -  
e^{- 2 \eps |t|}  \Delta E_D (p) \ .  }
\ee

The counter-term $\Delta E(p)$ is needed
for the convergence of the M{\o}ller operators for $\eps \to 0$,
which is obtained for $\Delta E_D (p) = \Delta E (p) $, eq.\,(5.6);
it has to cancel the photon contribution to the particle energy, 
which is of the second order in $e$ (Section 5).


\begin{Proposition} 1) The Hamiltonian $H$, eq.\,(2.4), is 
essentially self-adjoint  on the dense domain $ D $ of $\Cou$ 
functions $\psi(q, p) $    with values in $D(H_0)$;

\noindent 2) its adiabatic version for $\eps > 0$   
\be{ H^\eps(t) = h + H_0 + H_{I,\, r}^\eps (t) }\ee 
defines a family 
$U^\eps(t) = u(t)\,\U_0(t)\, \U^\eps(t)$, $\U_0(t) \eqq e^{- i H_0\,t}$,  
of unitary operators as the unique solution, 
leaving $D$ invariant, of 
\be{ i (d/ dt) \,U^\eps(t)\,\psi = H^\eps(t) \,
\uepst \,\psi, \,\,\,\,\,\forall \psi \in D \, ,}
\ee
given by (with $f(k;q,p)$ defined by eq.\,(2.6)) 
\be{ \U^\eps(t) =  
\exp{( - i \int_0^t  d s \,e^{- \eps |s|} \, H_I(s) )} 
\, e^{i \Phi^\eps_t - i \varphi^\eps_t} \,  ,} \ee
\be { H_I(s) = a(f_s) + a(f_s)^* \, ,\  
f_s(k; q, p) \eqq e^{-i |k| s}\, f(k; q_s(q, p), p_s(q, p)) ,}\ee 
\be{ \Phi^\eps_t \eqq \Phi^\eps_t(q,p) 
\eqq   i/2 \int_0^t
ds \int_0^s ds' \,e^{ - \eps(|s| + |s'|)} [\,H_I(s), \,H_I (s')\,] \ ,  }\ee
\be{ \varphi^\eps_t \eqq \varphi^\eps_{t}(q,p) 
\eqq -  \int_0^t d s \,
  e^{ - 2 \eps |s| } 
\Delta E(p_s(q,p))  \, ;}\ee

\noindent 3) the operators 
\be
{ H_D^\eps(t) \eqq  h_D (t) + H_0 +  H^\eps_{I D}(t) }
\ee
are hermitean on $D$ and the equation
\be{i (d/ dt) \uepsdt  = H_D^\eps(t)\,\uepsdt }\ee 
\noindent 
defines a family  $\uepsdt = u_D(t)\,\U_0(t)\,\U_D^\eps(t)$, 
of unitary operators as its unique solution leaving $D$ invariant, 
with $\U_D^\eps (0) = 1$, given by 
\be{ \U_D^\eps(t) = 
\exp{( - i \int_0^t  d s \,e^{- \eps |s|} \, 
H_{I D}(s) ) } \, e^{i \Phi^{D\,\eps}_t -i \varphi^{D\,\eps}_t} \, ,}  \ee
\be{ H_{I D}(s) = a(f_s^D) + a(f_s^D)^*, \,\,\,\,\,
f_s^D(k; p) \eqq e^{- i |k| s}\,f(k; vs, p) \ ,}\ee 
\be{ \Phi^{D\,\eps}_t \eqq   \Phi^{D\,\eps}_t (p) \eqq   
i/2 \int_0^t d s \int_0^s d s' e^{ - \eps(|s| + |s'|)} 
[\,H_{I D}(s), \,H_{I D} (s')\,] \, ,}
\ee
\be
{ \varphi^{D\,\eps}_t \eqq   \varphi^{D\,\eps}_t (p) \eqq   
-  \int_0^t ds \, e^{-2 \eps |s|} \, \Delta E (p)\ ; }
\ee

\noindent 
4) $\forall \eps > 0$, the following strong limits exist:
\be{  \limpm \uepst^* \, u_D(t) \, \U_0(t)  \equiv
W_{0 \pm}^\eps\, \om_\pm\, ,}
\ee
$$ \uepsdt^* u_D(t) \, \U_0(t) = \U^{\eps \, *}_D(t) \to_{t \to \pm \infty} $$
\be
\to 
\exp (i[ a(F_\pm^{D\,\eps}(p)) + a(F_\pm^{D\,\eps}(p))^*)]) \, 
\exp (- i \Phi^{D\,\eps}_\pm (p) + i \varphi^{D\,\eps}_\pm (p))  
\eqq W_{D \, \pm}^{\eps} \, ,
\ee
$$  F_\pm^{D\, \eps}(k, \lambda; p) = -i 
\sum_{i=1}^N J_\pm^{\eps}(k, \lambda, p_i)\, ,
$$
\be
{ J_\pm^{\eps}(k, \lambda, p) = 
\frac{ e }{(2 \pi)^{3/2} } \,
\frac{\epsilon (k, \l) \,\eta(k)\, v(p)}{(2 |k|)^{1/2} 
(|k| - v(p) k \mp i \eps)}\, , }
\ee
\be{ \limpm \uepst^* U_D^\eps(t)  = 
W_{0 \pm}^\eps \,\om_\pm \,W^{\eps \,*}_{D \pm} =
W_{0 \pm}^\eps (q,p)\,W^{\eps \,*}_{D \pm} (p_\pm) \,\om_\pm 
\eqq \Omega^\eps_\pm \, ;}
\ee
$ W_{0 \pm}^\eps (q,p), \,W^{\eps }_{D \pm} (p)$ are time ordered Weyl 
exponentials, acting on $\H_F$ and indexed by the particle variables $q,p$;  
they are given by eqs.\,(4.5), (4.11) with $ t = \pm \infty$, 
and similarly for $\Phi^{D\,\eps}_\pm$, $\varphi^{D\,\eps}_\pm $.
\end{Proposition}

\noindent
{\bf Remark.} Eqs.\,(4.5),\,(4.11) provide the existence and 
the explicit expression of the standard formula for $\U^\eps(t)$ 
and $\U_D^\eps(t)$ in terms of time ordered exponentials,
$$ \U^\eps(t) =  T( e^{-i\int_0^t ds \,  H_{int}^\eps (s) }) \, ,$$
where $T$ denotes the chronological ordering, according to the free photon 
dynamics and
$$H_{int}^\eps (s) \eqq \U_0^*(s) u^*(s) H_{I, \, r}^\eps (s) u(s) \U_0(s) \, ;$$
$$ \U_D^\eps(t) = T( e^{-i\int_0^t ds \,  H^\eps_{int \, D} (s) }) \, ,$$
$$H^\eps_{int \, D} (s) 
\eqq \U_0^*(s) u^*_D(s) H_{I, \, r}^\eps (s) u_D(s) \U_0(s) \, .$$

Apart from the mass renormalization counter-term, $\U_D^\eps(t)$ is 
the same operator used in \cite{KF} and \cite{JR} for the 
identification of the 
Dollard reference dynamics, with their asymptotic current taken as a function 
of our classical variables $p_i$.  

For the proof of Proposition 4.1 we need the following Lemma, 
proved in Appendix C.

\begin{Lemma} Let $f_\a(k), |k|^{-1/2}f_\a(k) \in L^2(d^3 k)$, $\a \in \Rbf$; 
if they are  differentiable  with respect to $\alpha$ in $L^2(d^3 k)$, then, 
\be{U(f_\a) \eqq e^{i(a(f_\a) + a(f_\a)^*)},}\ee
is strongly differentiable  on $D(H_0)$ and 
\be{ \frac{d U(f_\a)}{d \a} = [i (a(f'_\a) + a(f'_\a)^*) + C_\a ] \, U(f_\a)}\ee

\noindent 
with $f'_\a \equiv \partial_\a f_\a $, $ 
C_\a \eqq \ume \int d^3 k\, (f_\a \overline{f'_\a}  - 
\overline{f_\a}f'_\a ).$
Moreover, 
if $|k| f_\a \in L^2(d^3 k)$, then $U(f_\a)$ leaves $D(H_0)$ invariant.

\end{Lemma} 

\def \adie {a^\#}        \def \ltk {L^2(d^3 k)} \def \kum {|k|^{-1/2}}

\def \Fetk  {F_t^\eps(k; q, p)}
\def \Fet  {F_t^\eps(q, p)}
\def \FDetk  {F_t^{D\,\eps}(k;  p)}
\def  \FDet  {F_t^\eps(q, p)}

\vspace{1mm} 
\noindent \Pf (of Proposition 4.1). 

\noindent
1)-3). For fixed $q, p$, the argument of the exponential which defines 
$\U^\eps(t)$ is of the form $- i (a(F^\eps_t(q,\,p)) + a(F^\eps_t(q,\,p))^*)$, 
with 
\be{ F_t^\eps(k; q,\,p) = \int_0^t d s\,e^{- \eps |s|} f_s(k; q,\,p)
 \eqq \sum_i F^\eps_t (k, q_i , p_i ) \, .}\ee
\noindent A similar form holds for $\U_D^\eps(t)$, with $ f_s(k; q, p)$ 
replaced by $f^D_s(k; p)$. 
Both  $F_t^\eps(k; q,\,p)$  and $\FDetk$ satisfy the conditions of Lemma 4.2, 
with respect to $t, \,q,\, p$. 
Therefore, $\forall \eps \geq 0$ the right hand sides of 
eqs.\,(4.5), (4.11) define unitary operators 
$\U^\eps(t)$, $\U_D^\eps(t)$ in the Fock space $\H_F$, 
indexed by the particle coordinates $q,\,p$. 

\noindent Such unitary operators  
 leave $D(H_0)$ invariant and are strongly 
differentiable with respect to $t, \,q,\, p$, on $D(H_0) $.
Hence, in $\H$ the unitary operators 
$U^\eps(t)$, $U_D^\eps(t)$ leave $D$ invariant. 

\noindent Again by Lemma 4.2, $U^\eps(t)$, $U_D^\eps(t)$ are strongly 
differentiable  with respect to $t$ on $D$ and satisfy 
eqs.\,(4.4), (4.10), respectively.  
Hermiticity of $H^\eps(t)$ implies
$d/dt \, V^*(t) \, U^\eps(t) = 0 $ for any solution
$V(t)$ of eq.\,(4.4) leaving $D$ invariant and therefore 
uniqueness; the same for $H_D^\eps(t)$. 

\noindent For $\eps > 0$, this implies 2), 3). 
For $\eps = 0$, the uniqueness of the solution of eq.\,(4.4) implies 
that $U(t)$ is a one-parameter group; 
eq.\,(4.4) and the invariance of $D$ imply the self-adjointness of $H$. 

\noindent 
4). The left hand side of eq.\,(4.15) reads
$$
\U^{\eps\, *}(t)\, \U_0(t)^* \,u(t)^* \, u_D(t) \, \U_0(t) 
= \U^{\eps\, *}(t)\,u(t)^* u_D(t). 
$$
Using Proposition 3.1 and  eqs.\,(4.7),\,(4.8),\,(4.21), it converges to    
\be{e^{ i \int_0^{\pm \infty}  ds\, e^{ - \eps |s|} H_I(s)} \,
 e^{- i \Phi^{\eps}_{\pm \infty} + i \varphi^{\eps}_{\pm \infty} 
  \,\om_\pm }}\ee
$$ \equiv e^{ i (a(F_\pm^\eps(q,\,p)) + a(F_\pm^\eps(q,\,p))^*)} \,
 e^{- i \Phi^{\eps}_\pm + i \varphi^{\eps}_\pm } 
  \,\om_\pm \, .
$$
In fact,  

\noindent i) for $|t| \ra \infty$, $ F_t^\eps(k; q, p)$ 
converges in $\ltk$ uniformly in $ q, p$ on compact sets $K$, so that  
$\exp{ i\, (a(F_t^\eps(q,\,p)) + a(F_t^\eps(q,\,p))^*)} $ converges strongly on 
$ L^2(K, dq \,dp) \otimes D_{fin}$ and therefore everywhere;

\noindent 
ii) by similar arguments, $||f_s(k; q, p)||_{L^2(d^3 k)}$ is uniformly bounded 
in $s$ and in $q, p$ on compact  sets, so that 
\be{ < f_s, \,f_r > 
\eqq \int d^3 k\, (f_s\,f_r^* - f_s^* f_r)  }
\ee
is uniformly bounded in $s, r$ and in $q, p$ on compact sets. Then
$$ \Phi^\eps_t(q,p) =  i/2  \int_0^t d s \int_0^s d r\,e^{-\eps (|s| + |r|)} 
 \, < f_s, \,f_r > 
$$
converges for $|t| \ra \infty$. 

\noindent 
The left hand side of eq.\,(4.16) reads
\be{e^{ i \,(a(F_t^{D\,\eps}(p)) + a(F_t^{D\, \eps}(p))^*)}  
\,e^{- i \Phi^{D\,\eps}_t + i \varphi^{D\,\eps}_t }
\, , \ \ \  F_t^{D\, \eps}(k; p) \eqq \int_0^t d s 
\,e^{ - \eps |s| }\,f_s^D(k; p)}\ee
and converges as $|t| \ra \infty$, 
by the above argument applied to $f_s^D(k; p)$. 
An explicit calculation gives eqs.\,(4.16), (4.17) and the 
unitarity of $W_{D \pm}^\eps$, as in Lemma 4.2.  

\noindent The first equality in eq.\,(4.18) follows from eqs.\,(4.15),\,(4.16)  
and the unitarity of $W_{D \pm}^\eps$, which implies 
the convergence of the adjoint of eq.\,(4.16). The second
equality follows from eq.\,(3.12). $\endproof $




\section{Removal of the adiabatic switching}

In this Section we perform the limit $\eps \to 0 $; 
the crucial ingredient is the use of the Dollard reference 
dynamics $ U_D^\eps(t)$, but, as anticipated in Sect. 2, 
the introduction of suitable counter-terms will be required.
We have to consider the behavior of the operators $\Omega^\eps_\pm$, 
eq.\,(4.18), as $\eps \to 0$.
By using eqs.\,(4.21),\,(4.22),\,(4.16), one has 
$$ \Omega^\eps_\pm = 
e^{ i (a(F_\pm^\eps(q,\,p)) + a(F_\pm^\eps(q,\,p))^*)} \, 
e^{-i( a(F_\pm^{D\,\eps}(p_\pm)) + a(F_\pm^{D\,\eps}(p_\pm))^*)} \times $$ 
\be
{e^{i (-  \Phi^\eps_\pm +  \Phi^{D\,\eps}_\pm(p_\pm))} 
e^{i (\varphi^\eps_\pm - \varphi^{D\,\eps}_\pm(p_\pm))} 
\, \om_\pm \equiv
 \Omega (\Delta F_\pm^\eps) 
\ e^{- i \Delta \Phi_\pm^\eps + i \Delta \varphi_\pm^\eps} 
\  \om_\pm \  ,}
\ee
$\Omega (\Delta F_\pm^\eps) $ denoting the product of the two
Weyl exponentials in the l.h.s.. 

The convergence of the term $ \Omega (\Delta F_\pm^\eps) $
amounts to the cancellation of the infrared divergences associated
to infinite photon emission, thanks to the Dollard subtraction given by
the coherent factors, eqs.\,(4.16),\,(4.17).

For the convergence of the phases, we shall use the fact that
$\Phi^\eps_\pm$ and $\Phi^{D\,\eps}_\pm$ involve the commutators
$$ [ A_i(x)\, , \, A_j(y)] 
\equiv i D_{i\,j} (x-y) \ , \ \ \ x,y \in \dR^4 \ , \ \ i,j = 1,2,3 \, . 
$$
The fields $A_i (x)$ are free because their time dependence is given
by the interaction representation and one has
 ($ x^2 = x_0^2 -  {\mathbf x}^2 $)
\be {
D_{i\,j}(x) 
= \delta_{i\,j} D(x) +  \partial_i \partial_j 
(\SI (x_0) \, \theta(x^2) + x_0/| {\mathbf x} | \, \theta(-x^2) ) / 4 \pi \ ,}
\ee
with $D(x)$ 
the standard commutator function.
$ D_{i\,j}(x)$ 
has spacelike support and it is homogeneous of degree $-2$.

We denote by $ D_{i\,j}^\eta (x)$  
the double convolution in the space variables 
$$  D_{i\,j}^\eta (x) 
\equiv  \int  d^3 \xi \, d^3 \eta \,
D_{i\,j} (x - (\xi - \eta)) \,
\tilde \eta (\xi) \,  \tilde \eta (\eta) \, . 
$$
Considering the case of $t >0$ , we have to control the $\eps \to 0$
limit of   
\be{ \Phi^\eps_{+} (q,p) = \ume  
\sum_{m,n = 1}^N 
\int_0^\infty G^\eps_{m\,n} (x_0,q,p) 
\, dx_0 
\eqq \ume \sum_{m,n = 1}^N (\Phi^\eps_{+})_{mn} \ ,}
\ee
where, omitting as before the vector notation for $q$ and $v$,
$$
G^\eps_{m\,n} (x_0,q,p) 
\equiv 
 e^2 \int_{0 \leq y_0 \leq x_0} d^3x \, d^4y \, e^{-\eps(x_0 + y_0)} \,
v_m(x_0) \, v_n(y_0) \,  D^\eta (x - y) \times
$$ 
\be {\delta ({\mathbf x} - q_m(x_0)) \, 
\delta ({\mathbf y} - q_n(y_0)) \, , }
\ee
with  $v_n(x_0)$, $q_n(x_0)$   
given by the solution at time $x_0$ of the equations
of motion with initial data $(q,p)$ and  
$ v_m \, v_n \,  D^\eta \equiv \sum_{i\,j =1}^3
v_{m\,i} \, v_{n\,j} \,  D_{i\,j}^\eta $.

\noindent 
$\Phi^{D\,\eps}_+ (p_+)$ is given by eq.\,(5.3) with 
$ G^{\eps}_{m\,n} $ replaced by $ G^{D\,\eps}_{m\,n}$, 
$$ G^{D\,\eps}_{m\,n} (x_0, q,p) \eqq 
 e^2 \int_{0 \leq y_0 \leq x_0} d^3x \, d^4y \, e^{-\eps(x_0 + y_0)} \,  
v_{m \, +} \, v_{n \, +} \,  D^\eta (x - y) \times
$$
\be { \delta ({\mathbf x} - v_{m \, +} x_0) \,
    \delta ({\mathbf y} - v_{n \, +} y_0) 
\eqq  G^\eps (x_0, p_{m \, +}, p_{n \, +})    }
\ee
(reproducing eq.\,(S4.21) in \cite{JR}).

The convergence of the off-diagonal terms in the phase difference 
$\Phi^\eps_\pm - \Phi^{D\,\eps}_\pm(p_\pm)$, 
eq.\,(5.7) below  with $n \neq m$,
amounts to the Dollard cancellation of the Lienard-Wiechert corrections to
the Coulomb phases arising from ``photon exchanges''.
The corresponding diagonal terms, $n=m$, are logarithmically divergent, 
eq.\,(5.7) with $n=m$, even if
their $1/\eps$ divergent terms are canceled by Dollard's subtraction; 
they correspond to a logarithmically divergent 
``mass renormalization'' effect
produced by the Coulomb asymptotic distortion of the trajectories
(a point which is not discussed in \cite{JR}).
The problem is solved by the introduction of the same
``mass renormalization'' counter-term $\Delta E(p)$ in  $H_I$ and in
$H_D$,
\be 
{  \Delta E (p) = \sum_n \delta E (p_n) \eqq
 \ume  \sum_n   e^2  \int_{y_0 \leq 0} d^4y \, v_n^2 D^\eta (-y)
  \delta ( \mathbf {y} - v_n y_0) \, ; }
\ee
it corresponds to the linearization, with parameters $p_n$,
of the particle trajectories in eq.\,(5.4),
$q_n(y_0) = q_n(x_0) + (y_0 - x_0) v_n$, $v_n=v(p_n)$.
We denote by $(\varphi^\eps_\pm)_n $ the contribution of  
$\delta E (p_n) $ to $\varphi^\eps_\pm $, see eq.\,(4.8).

Actually, the introduction of the above counterterm in the Dollard 
dynamics cancels 
the divergence of 
the Dollard phases $(\Phi^{D\,\eps}_\pm)_{nn}$, and therefore 
the convergence of the diagonal terms reduces to the convergence of
$(\Phi^\eps_\pm)_{nn} - (\varphi^\eps_\pm)_n$, 
i.e. to the effect of the counterterm.

In the following, for simplicity, in the non-relativistic case, 
we shall write $\H$ for $\H_{nr}$ and 
$ \Gamma_\pm $ for $ \gamma_\pm^{-1} (\Gamma_{nr}) $, 
$\Gamma_{nr} $ the set of non-relativistic particle configurations, 
eq.\,(2.8). We also introduce  
$$\H_\pm \eqq \omega_\pm L^2(\Gamma) \times \H_F = 
L^2(\Gamma_\pm) \times \H_F \, .$$

\subsection{The M{\o}ller operators}
\begin{Proposition} As $\eps \to 0$,

\noindent
1)  $ \Omega (\Delta F_\pm^\eps)  $, eq.\,(5.1)  
converge strongly, to unitary operators, on $\H_\pm$;

\noindent
2) for almost all $(q,p)$ in $\Gamma_\pm$, 
\be 
{( \Phi^\eps_\pm - \Phi^{D\,\eps}_\pm(p_\pm))_{mn} = \delta_{m\, n}
  {R(p_\pm , \delta p_\pm)} \, \ln \eps  + O(1) \ , }
\ee
with $R(p_\pm , \delta p_\pm) $ defined below, see eq.\,(5.11);

\noindent
3) with the choice of the counterterm given by eq.\,(5.6), we have
$$
  (\Phi^{\eps}_\pm)_{nn} -  (\varphi^{\eps}_\pm)_n  = O(1) \,  , 
\ a.e. \ in \ \Gamma_\pm 
$$
and 
$$
  (\Phi^{D\,\eps}_\pm (p_\pm))_{nn} -  (\varphi^{D\,\eps}_\pm (p_\pm))_n  
 \eqq \delta \varphi_\pm^{D \, \eps} (p_{n \, \pm})   =  O(1) \, ;
$$
the phase $\delta \varphi_\pm^{D \, \eps} (p_{n \, \pm}) $
 vanishes after the redefinition 
$$ 
\U_D^\eps \to \U_D^\eps \, e^{i \sum_i \delta \varphi_\pm^{D \, \eps} (p_i)}
$$ 
for $ \pm t > 0$, which shall be understood in the following.

\noindent
Therefore,  
with the above choice of the counterterms, 
the following strong limits exist:
$$ \lim_{\eps \to 0} 
\Omega_\pm^\eps \eqq \Omega_\pm = W_\pm\, \omega_\pm \ \ \
 on \ \H  \ , \ \ \  W_\pm \ unitary \  in \ 
 \H_\pm =  \omega_\pm \, \H  \ , $$
\be {
 W_\pm = \lim_{\eps \to 0} W_{0 \pm}^\eps (q,p) \, W^{\eps \,*}_{D \pm} (p_\pm) \ \ 
 \ on \ \H_\pm \ .   }
\ee

\end{Proposition}

\noindent
\Pf \,1). Omitting the dependence on polarization vectors, we put 
$$ \Delta F_\pm^\eps (k) 
\equiv F_\pm^\eps (k;q,p) - F_\pm^{D\,\eps} (k;p_\pm(q,p)) $$   
and we have (with the notation of Lemma 4.2 and of eq.\,(4.23))
$$  \Omega (\Delta F_\pm^\eps) ) = U( \Delta F_\pm^\eps )
\, e^{\frac{1}{2} < F_\pm^\eps  \, , \, F_\pm^{D\,\eps} >} \ . $$  
We shall prove convergence, in $L^2(d^3 k)$,  
almost everywhere in $\Gamma_\pm$,
of  both i) $\Delta F_\pm^\eps (k,q,p)$ and 
ii) $|k|^{-1/2} \, \Delta F_\pm^\eps (k,q,p)$.
i) implies strong convergence of 
$U (\Delta F_\pm^\eps)$ in Fock space, for fixed $q,p$, and
their strong convergence, as multiplication operators in $q,p$, 
on $  \H_\pm  $, by a Lebesgue dominated convergence 
argument, to unitary operators.

\noindent
Moreover, by eq.\,(4.17), for fixed $p$ in 
$ \Gamma_\pm $, 
$|k|^{1/2} \, F_\pm^{D\,\eps} (k, \lambda ; p)$
converges in $L^2(d^3 k)$ as $\eps \to 0$.
Then, since
$$ <F_\pm^\eps  \, , \, F_\pm^{D\,\eps}> \, =  \,
<\Delta F_\pm^\eps  \, , \, F_\pm^{D\,\eps}>  \ , $$  
ii) implies convergence of the phase factors, pointwise and therefore
strongly on $  \H_\pm $.

\noindent 
It is enough to prove ii), which 
implies i) thanks to the ultraviolet cutoff $\eta(k)$.
To this purpose, we note that eqs.\,(3.13),\,(3.18) imply that, 
for almost all $(q,p)$ in  $ \Gamma_\pm $,
for large $|s|$, 
$$ |k|s - k \, q_{i\,s} \eqq   (|k| - k \, v_{i\,\pm} ) \, s'(s, k/|k|) 
\equiv \theta(k/|k|) \, |k| \, s' $$   
defines a function 
$s'(s, k/|k|)\equiv s - \Delta s $, satisfying
$J(s) \equiv \partial \Delta s/ \partial s = O (s^{-1})$ and
$\Delta s = O (\ln\,|s|)$,
uniformly in $k/|k|$;  
$s(s')$ will denote the inverse function,
for large $s'$, at fixed $k/|k|$. 
Then, considering for simplicity positive times, apart
from an integral over a finite time interval, 
$0<s< c$, giving rise to a convergent term,  
$|k|^{-1/2} \, \Delta F_{+}^\eps $ 
may be written as $\eta(k)/|k|$ times
$$ \int_c^{\infty} ds \, e^{- \eps s}  e^{-i \, |k|\,  s} \, ( e^{ik q_s(q,\, p)} \, v_s - 
e^{ik  v_{+}(q,\, p) s } \, v_{+} ) = $$
$$ = \int_c^{\infty} ds \,  e^{-\eps s} \, 
( e^{-i \theta \, |k| \, s'} - e^{-i \theta \, |k| \, s} )\, v_{+} + 
\int_c^{\infty} ds' \, e^{- \eps s(s')} e^{-i \theta \, |k| \, s'} \, g(s', k/|k|) \ , $$
with $g(s', k/|k|) $  of order $1/s'$ and therefore in $L^2(ds')$,
with norm boun\-ded uniformly in $k/|k|$.
Since, for bounded momenta, $\theta$ is bounded away from $0$,
the last term converges in $L^2(d|k|)$, uniformly in $k/|k|$, as $\eps \to 0$.
Therefore, it gives a contribution to  $|k|^{-1/2} \, \Delta F_{+}^\eps $ 
which converges in $L^2(d^3 k)$.
By an obvious change of variables, and omitting as before the integration 
over a finite interval, the first term can be written 
$$  \int_c^{\infty} ds \, e^{-i \theta \, |k| \, s}  \,  
e^{- \eps s} (e^{- \eps \Delta s }/(1+J) - 1) \, p_{+} . $$ 
Therefore, it is the Fourier transform of a function which converges, as 
$\eps \to 0$, in $L^2(ds)$, since $1/(1+J) = 1 + O(s^{-1}) $ and,
for $s \geq 1$,
$$  e^{ \eps \ln s} - 1  \leq \eps \ln s  \, e^{ \eps \ln s} \ . $$ 
Hence, its contribution to  $|k|^{-1/2} \, \Delta F_{+}^\eps $ 
converges in $L^2(d^3 k)$, with convergence rate 
$O(\eps^{1/2 - \delta})$, $\forall \delta >0$ . 
Similarly for $\Delta F_{-}^\eps $. 

\noindent
2). First, we consider the terms corresponding to $m \neq  n$, with non
collinear $v_{m \, +}$, $v_{n \, +}$, a condition which holds almost 
everywhere in $  \Gamma_+ $, by eq.\,(3.12).
For their contribution to $\Delta \Phi_{+}^\eps $ 
we exploit Lemma 5.2 below.
In fact, as a consequence of eqs.\,(3.13),\,(3.18), the particle 
trajectories satisfy eqs.\,(5.12)-(5.14), 
for any $\alpha < 1 $, a.e. in $ \Gamma_+ $.

Then, for $m \neq n$, uniformly in $\eps$, 
$$ G^\eps_{m\,n} (s) - G^{D\,\eps}_{m\,n} (s) \leq O (s^{-(1+\alpha)} $$
and the contribution to $\Delta \Phi_{+}^\eps $ is convergent. 
with rate $O(\eps^\alpha)$. 

\noindent
For $n = m$, we compute the logarithmic divergences,
which arise from the subleading terms in the asymptotic estimates
of the trajectories.

\noindent 
In this case, for $q,p$ in $ \Gamma_+ $,
the velocities $| v_i | $ have a bound less than 
$1$ and therefore, by the support properties of $D$ and $\eta$, 
$x_0 - y_0$ is bounded uniformly in $x_0$, i.e., $x_0 - y_0 < T $, 
in the integration in eq.\,(5.4).
The diagonal term $G^\eps_{nn}$ is a functional $\G^\eps $
of the $n$-th particle trajectory, 
$\{ q_n(\tau), \, p_n(\tau) \}  \eqq  \{ (q_{n \, \tau}(q,p), \,   
 p_{n \, \tau}(q,p)), \, \tau \in \dR \}$,  
\be G^\eps_{nn} (x_0, q,p) 
=  \G^\eps (x_0; \{ q_n(\tau), \, p_n(\tau) \})  \  ;
\ee
the expression for the corresponding Dollard term is 
given by the Dollard trajectories,
$ G^{D \, \eps}_{nn} (x_0, q,p) =
  \G^\eps (x_0;\{ v_{n \, +} \tau, \, p_{n \, +} \} ) $.
We introduce 
$$
 G^\eps_{n \, +} (x_0, q,p) 
\equiv  \G^\eps (x_0; \{ v_{n \, +}\tau + 
v'(p_{n \, +}) \, \delta p_{n \, +} \ln \tau, \,
p_{n \, +} + \delta p_{n \, +} / \tau \} )  \ ,
$$
$v'(p) \eqq \partial v(p) / \partial p$, 
which satisfies, a.e. in $ \Gamma_+ $,
\be | G^\eps_{nn} (x_0, q,p) - G^\eps_{n \, +} (x_0, q,p) | 
\leq C(q,p) \ln  |x_0| / x_0^2 \ \ ,
\ee
uniformly in $\eps$.
In fact, by eq.\,(3.13), omitting the index $n$, 
$$  q(x_0) - q(y_0) - v_+ (x_0 - y_0) - 
 v'(p_{+}) \, \delta p_+ \ln (x_0/y_0) = $$
$$  \int_{y_0}^{x_0} ds \, 
(v(s) - v_+ - v'(p_{+}) \delta p_+ / s)  
=  O(\ln x_0 / x_0^2) (x_0 - y_0) = O(\ln x_0 / x_0^2) T  ; $$
since $D^\eta(x)$ is a $C^\infty$ function of $\mathbf x$
and $p_n(t) $ satisfies the estimate (3.13), eq.\,(5.10) follows, uniformly
in $\eps$ since $x_0 - y_0 \leq T$. 
In conclusion, the contribution to $ \Delta \Phi_+^\eps $ is  
$$ 
(\Delta \Phi_+^\eps)_{nn} = 
\int dx_0 \, (G^\eps_{nn} (x_0, q,p) - G^{D \, \eps}_{nn} (x_0, q,p))  =
$$
\be \int dx_0 \, (G^\eps_{n \, +} (x_0, q,p) - G^{D \, \eps}_{nn} (x_0, q,p)) 
 + O(1) \ .
\ee    
$G^\eps_{n \, +}  - G^{D \, \eps}_{nn} $ 
is a function of $p_{n \, +} , \delta p_{n \, +}$ and $x_0$; 
by a Taylor expansion in $\tau $ around $\tau = x_0$, it
is of the form  
$ ((R(p_{n\, +} , \delta p_{n \, +}) + O(\eps)) \, x_0^{-1} \, e^{-2\eps x_0}
+ O(x_0^{-2})$ and 2) follows. 
Similarly for $(\Delta \Phi_-^\eps)_{nn}\, $. 

\noindent
3). $ (\Phi^{D\,\eps}_+ (p_{+}))_{nn}$ is given by eq.\,(5.5) 
and $ (\varphi^{D\,\eps}_+ (p))_{n}$, see eqs.\,(4.14), (5.6), is given by
the same expression, without the factor  
$ e^{-\eps (x_0 + y_0)} $ and the restriction $ 0 \leq y_0 $. 
Hence, they only depend on $p_{n \, +} $
and their difference converges as $\eps \to 0$.

\noindent
With a change of variables, $\mathbf{x'} \eqq \mathbf{x} - q_n(x_0)$,
$\mathbf{y'} \eqq \mathbf{y} - q_n(x_0)$, $y'_0 = y_0 - x_0$
in eq.\,(5.4), 
putting $ G_{nn}^\eps = \hat G_{nn}^\eps  \, e^{- 2 \eps x_0} $,  
one has, for $(q,p)$ in $ \Gamma_+ $, 
$$
\ume  \hat G_{nn}^\eps (x_0,q,p) 
- \delta E (p_{n \, x_0} (q,p))  =
-  \ume e^2 
\int_{y'_0 \leq 0} d^4y' \, D^\eta (-y') \, v_n(x_0)  \times
$$
$$[ v_n(x_0 + y'_0) \, \delta (\mathbf{y'} - q_n(x_0 + y'_0) + q_n(x_0))
\, e^{\eps \, y_0'}
-  v_n(x_0) \, \delta (\mathbf{y'} - v_n(x_0) \, y'_0) ]
$$
As before, $|y'_0| \leq T$, and therefore
$ v_n(x_0 + y'_0) = v_n(x_0) + O (x_0^{-2})$, 
$ q_n(x_0 + y'_0) + q_n(x_0) = v_n(x_0) \, y'_0/m + O (x_0^{-2}) $.
Hence, 
$ \hat G_{nn}^\eps (x_0,q,p) - \delta E (p_{n\, x_0} (q,p)) \in L^1(dx_0) $
uniformly in $\eps$
and $(\Phi^\eps_+)_{nn} - (\varphi^\eps_+)_n $ converge for $\eps \to 0$,
with rate $O(\eps \ln \eps)$.

\noindent Convergence of the regularized M{\o}ller operators follows, with 
rate of convergence  $O(\eps^{1/2 - \delta})$, $\forall \delta >0$, on 
$\H_+ = L^2(\Gamma_+) \times \H_F $; their limit,
$W_+$,  is a product of phases and Weyl operators, acting as
multiplication operators on $ L^2(\Gamma_+) $, and therefore 
unitary operators in $\H_+$.  
The same applies for $t \to -\infty$. $ \ \ \endproof $

\begin{Lemma}
Let $q_n(x_0), v_n(x_0) \in C^1(\dR)$ satisfy, for $x_0 \to \infty$,
\be {|\dot q_n (x_0)| < 1- \delta  \ , }
\ee
\be
  q_n(x_0)/x_0  = v_{n \, +} + O(x_0^{-\alpha}) \ , 
\ee
\be
p_n(x_0) = p_{n \, +} + O(x_0^{-\alpha}) \ , 
\ee
with $\delta, \alpha > 0$. 
Then, for $v_{m \, +}, v_{n \, +}$ non collinear, $ \forall \varepsilon \geq 0$, 
denoting $ j_m (x) \equiv  v_m(x_0) \delta(\mathbf{x} -q_m(x_0))  $,
$ j_{m \, +} (x) \equiv  v_{m \, +} \delta(\mathbf{x} - v_{m \, +} x_0)  $,
omitting the vector notation as in  
eqs.\,(5.4),\,(5.5), 
$$
\int_{0<y_0<x_0}  d^3x \, d^4y \, e^{- \varepsilon  y_0}  \,
j_m(x) \,  D^\eta (x-y) \, j_n(y) =
$$
\be
\int_{0<y_0<x_0}  d^3x \, d^4y \, e^{- \varepsilon  y_0)} \,   
j_{m \, +} (x) \,  D^\eta (x-y) \, j_{n \, +}(y) \,  
+ \, O(x_0^{-1-\alpha}) \, ,
\ee
uniformly in $\varepsilon$. 
By homogeneity of $D_{i\,j}$, for $\varepsilon = 0$, 
the first term in the r. h. s. is of the form 
$ v_{n \, +} v_{m \, +} C(v_{m \, +}, v_{n \, +}) \, x_0^{-1}$.
\end{Lemma}

\noindent 
\Pf The l.h.s. of eq.\,(5.15) 
is well defined, for all $x_0$, thanks to the 
regularization given by $\tilde\eta$ and, by a change of variables,
$\tau \equiv y_0/x_0$, $\mathbf{x'} \equiv \mathbf{x} / x_0$, 
$\mathbf{y'} \equiv \mathbf{y} / x_0$, 
it becomes
$$
\frac{1}{x_0} \int_{0<\tau<1} d\tau\, d^3x'\, d^3y'\; v_m(x_0) \, 
v_n(\tau x_0) \, e^{- \varepsilon x_0 \tau)} 
$$
\be
D^{\tilde\eta_{x_0}}(\mathbf{x' - y'}, 1-\tau) \,
\delta(\mathbf{x'}- q_m(x_0)/x_0) \, 
\delta(\mathbf{y'}- q_n(\tau x_0)/x_0) \ , 
\ee
where $\tilde\eta_{x_0}(\xi) \equiv \tilde\eta( x_0 \, \xi) x_0^3$ and the 
homogeneity of $D_{i\,j}$ has been used.
For large $x_0$, the integrand of eq.\,(5.16) vanishes for
$\tau < \delta/2$. In fact, 
$D^{\tilde\eta_{x_0}}$ 
has spacelike support, apart from a correction of order
$x_0^{-1}$; moreover, $\tau < \delta/2$ and eq.\,(5.12) imply
$$ | q_m(x_0) - q_n(\tau x_0)| / x_0  \leq   
x_0^{-1} (| q_m(0) | + | q_n(0) |)  + (1-\delta)(1+\tau)   < 1 -\tau \ ,
$$
the last inequality following, for large $x_0$,  from
$$ (1-\tau)/(1+\tau) \geq (1-\tau)^2 > 1 -\delta \ . $$
Therefore, by eq.\,(5.13),
$ q_n(\tau x_0)/x_0 = \tau v_{n \, +} + O(x_0^{-\alpha}) $
in eq.\,(5.15). 
Moreover, for non collinear asymptotic velocities and
$x_0$ large, the support of integrand in eq.\,(5.16) excludes a 
neighbourhood of $x' - y' =0$.

\noindent
Outside a neighborhood of $\mathbf{x} = 0$, the 
second term in the representation  of $ D_{i\,j}(x)$,
eq.\,(5.2), is a bounded function with spacelike support and
bounded derivatives inside the spacelike region,
and the same applies to its convolution with $\tilde\eta_{x_0}$, 
with bounds uniform in $x_0$, apart from the addition of a uniformly bounded 
function with support within a distance of order $x_0^{-1}$ from the light cone; 
therefore, one may replace $ q_m(x_0)/x_0 \mapsto \tau v_{m \, +} $ and
$ q_n(\tau x_0)/x_0 \mapsto \tau v_{n \, +} $ in eq.\,(5.15), with an error of
order $x_0^{-\alpha}$. 

\noindent
The first term, $\delta_{i\,j} D(x)$, only involves $\delta$ functions and may
therefore be treated explicitely; 
the result follows, for $v_{m \, +}$, $v_{n \, +}$ non collinear, from  
eq.\,(5.13) and eq.\,(5.14),
the convolution with $\eta_{x_0}$ giving rise to
corrections of order $x_0^{-2}$.  $ \ \ \endproof $



\subsection{The asymptotic dynamics}

As shown in \cite{MS}, any Dollard reference 
dynamics allowing for the existence 
of M{\o}ller operators defines asymptotic dynamics, for $t \to \pm\infty$, 
$U_\pm(t)$, which need not to coincide with the free dynamics, but are 
always one parameter groups, satisfying the M{\o}ller intertwining 
relations. In presence of an adiabatic procedure, the latter property 
involves the recovery of the  dynamics from its adiabatic 
regularization \cite{MS}.

Therefore, the next step in the analysis of the model 
is the determination of the asymptotic dynamics,
and the verification of the intertwining relations. 

The resulting asymptotic dynamics, $U_+(t) = U_-(t) \eqq U_{as}(t)$
is uniquely determined by the Dollard dynamics $U_D(t)$ but cannot
be identified with it, as implicit in Rohrlich and Zwanziger notions
of asymptotic fields and dynamics.
 \goodbreak

\begin{Proposition} 
With  the 
counterterm given by eq.\,(5.6), one has

\noindent
1) the existence of the following strong limits
\be { U_\pm (s) \eqq \lim_{\eps \to 0} \lim_{t \to \pm \infty} 
U_D^{\eps\, *}(t) U_D^{\eps}(t+s) \, ,}
\ee
which define the asymptotic dynamics
$$  U_+(s) = U_{-}(s) = u_0(s) \, \alpha_{as} (\U_0 (s)) \eqq U_{as}(s) \, , $$
with $\alpha_{as} $ the coherent automorphism of the photon algebra 
\be
 \alpha_{as} (a^*(k,\lambda)) = a^*(k,\lambda) + J(k,\lambda; p) \ ,
\ee
$ J(k,\lambda; p) = \sum_i J (k, \lambda, p_i)$,  
$ J = J_\pm^{\eps = 0} $,  
see eq.(4.17); 
in the non-re\-la\-ti\-vi\-stic case, $U_{as}(s) $ 
leaves $\H_{nr} $ invariant;

\noindent
2) the recovering of $U(t)$ from the regularized dynamics $U^\eps (t)$:
\be 
{ \lim_{\eps \to 0} \lim_{t \to \pm \infty} U^{\eps\, *}(t) U^{\eps}(t+s) \eqq
  \lim_{\eps \to 0} \tilde {U}^\eps (s) = U(s) \, , }
\ee
all the limits being strong;

\noindent 
3) the interpolation formula
\be{ U(t) \, \Omega_\pm = \Omega_\pm \, U_{as} (t) \, ; }\ee

\noindent
4) covariance under space translations,
\be{ \P \, \Omega_\pm = \Omega_\pm  \, \P_{as}\ \ \ \ 
 \P_{as} \eqq i \sum_i \partial/ \partial q_i  + \alpha_{as} (P_{ph}) \, .}\ee

\end{Proposition}
\noindent 
\Pf\, 1). By definition,
$$
U_D^{\eps\, *}(t) U_D^{\eps}(t+s) = 
 \U_D^{\eps\, *}(t) \, \U_0^* (t) \, u_D^*(t) \, 
 u_D(t+s) \, \U_0(t+s)  \, \U_D^{\eps}(t+s) \, ;
$$
by eq.\,(4.16), $[u_D, \, \U_0] = 0$ and eq.\,(3.10), the above expression 
converges strongly,
for $t \to \pm \infty$, to
$$
W^\eps_{D \, \pm} \, u_0(s) \, \U_0(s) \, W^{\eps \, *}_{D \, \pm}
\to_{\eps \to 0}   u_0(s) \, e^{-i H_{0}(J)s} \eqq U_{as}(s) \, ,
$$
\be
H_{0}(J) 
\eqq \sum_\lambda \int d^3k \,|\kbf| 
(a^*(k,\lambda) + J(k,\lambda;p) )
\, (a(k,\lambda) + J(k,\lambda;p) ) \ ;
\ee
$U_{as}(s)$ leaves $\H_{nr} $ invariant since so does $u_0$
and $H_{0}(J)$ acts as a 
multiplication operator on the particle space.\goodbreak 

\noindent
2). From eq.\,(4.5) one has (for large positive $t$)
$$
U^{\eps\, *}(t) U^{\eps}(t+s) = e^{- i \Phi^\eps_t + i \varphi^\eps_t} 
e^{ i \int_0^t  d s' \,e^{- \eps s'} \, H_I(s') )} \times
$$
$$
u(s) \, \U_0(s)
e^{ - i \int_0^{t+s}  ds' \,e^{- \eps s'} \, H_I(s') )} 
e^{ i \Phi^\eps_{t+s} - i \varphi^\eps_{t+s}} =
$$
$$
u(s) \, \U_0(s) e^{ i \int_s^{t+s}  d s' \,e^{- \eps s'} \, H_I(s') \, e^{\eps s}} 
e^{ - i \int_0^{t+s}  ds' \,e^{- \eps s'} \, H_I(s') )} \times
$$
$$
e^{( - i \Phi^\eps_{s,t+s} + i \varphi^\eps_{s,t+s}) e^{2 \eps s}} 
e^{ i \Phi^\eps_{t+s} - i \varphi^\eps_{t+s}} \ .
$$
Now, by Proposition 5.1, both the phase factors on the r.h.s. converge as 
$t \to \infty$ and $\eps \to 0$, as multiplication operators on  
$\Omega_+ ( \H )$, which coincides with $\H_+$ 
since $\Omega_+ = W_+ \omega_+$, (eq.\,(5.8)), 
with $ W_+$ unitary operators in $\H_F$, indexed by $q,p$.
Therefore, the factor $e^{2 \eps s}$ can be substituted by $1$ and
$$ \varphi^\eps_{s,t+s} - \varphi^\eps_{t+s} \to - \varphi_s \ , 
$$
\be 
{  - \Phi^\eps_{s,t+s} + \Phi^\eps_{t+s} \to \Phi_s + i/2 
\int_s^\infty dr_1 \int_0^s dr_2 \, [H_I(r_1), \, H_I(r_2)] }
\ee
(with a compact integration range, by locality).
The product of the exponentials involving $H_I$ 
can be written as the exponential of
$$
- i \int_0^{s}  ds' \,e^{-\eps s'} \, H_I(s')  
- i \int_s^{t+s}  ds' \,e^{-\eps s'} \, H_I(s') (1- e^{\eps s}) +
$$
$$
 \ume \int_s^{t+s} dr_1 \int_0^{t+s} dr_2 \, 
[H_I(r_1), \, H_I(r_2)] \, e^{- \eps (r_1 + r_2)} e^{\eps s}  \ .
$$
In the limit
$t \to \infty$, $\eps \to 0$, as a consequence of the antisymmetry 
of the integrand, the last term exactly cancels the last term on 
the r.h.s. of eq.(5.23). With the notation of eq.\,(4.19), 
the exponential of the first two terms is of the form
$ U(G^\eps(s,t))$, with 
$$
G^\eps(s,t) \eqq F^\eps_s 
+ (F^\eps_{t+s} - F^\eps_s)(1-e^{\eps s}) \ .
$$
For $t \to \infty$, 
with the notation of eqs.\,(4.17),\,(4.22),\,(5.1), one has that  
$ F^\eps_{t+s}  \to F^\eps_+ = F^{D \,\eps}_+ + \Delta F^\eps_+ \ , $
a.e. in $\Gamma_+$; for $\eps \to 0$,
$ \Delta F^\eps_+ $ converges in $L^2 (d^3k)$ and
$ || F^{D \eps}_+ ||_{L^2 (d^3k)}  $  is bounded by $O(\ln \eps)$,
uniformly in $q,p \in \Gamma_+$, apart from sets of 
arbitrarily small measure. 
Therefore, for $t \to \infty$ and then $\eps \to 0$, 
$ G^\eps(s,t) (k,q,p)$ converges to $ F^0_s$ 
in $L^2 (d^3 k)$, uniformly  in $q,p \in \Gamma_+$, apart from sets of 
arbitrarily small measure.
Then, strongly on $e^{\Phi_s - \varphi_s} \, \Omega_+ (\H_{nr} )$,
$$  
U(G^\eps(s,t)) \to U(F^0_s) = \exp {(-i \int_0^s ds \, H_I(s'))} \, ;
$$
(the phases and the operator $ W_+$ act as multiplication operators
in the variables $q,p$ and therefore leave the support of  
$\psi(q,p)\in \H_F $, $(q,p) \Gamma_+$ invariant).
The same applies for $t \to -\infty$ and eq.\,(5.19) follows.

\noindent
3). By using eqs.\,(5,19),\,(5.17), one has
$$  
U(s) \, \Omega_\pm = 
 \lim_{\eps \to 0} \lim_{t \to \pm \infty} U^{\eps\, *}(t) \, U^{\eps}(t+s) \,
 U^{\eps \, *}(t+s) \, U_D^{\eps}(t+s)  $$
$$  
 = \lim_{\eps \to 0} \lim_{t \to \pm \infty} U^{\eps\, *}(t) \, U_D^{\eps}(t) \,
   U_D^{\eps\, *}(t) \, U_D^{\eps}(t+s) = \Omega_\pm \, U_{as} (t) \, . $$

\noindent
4). By eqs.\,(5.8),\,(3.12), 
$$ \Omega_\pm = \lim_{\eps \to 0} W_{0 \pm}^\eps (q,p) \, \omega_\pm \,
  W^{\eps \,*}_{D \pm} (p)  $$
and, since the space translations $T(a)$ commute with $\omega_\pm$
and $W^{\eps \,*}_{0 \pm} (p) $,
$$ T(a) \, \Omega_\pm  
= \lim_{\eps \to 0} T(a) \, 
W_{0 \pm}^\eps (q,p) \, \omega_\pm \, W^{\eps \,*}_{D \pm} (p) $$
$$
 = \Omega_\pm \, \lim_{\eps \to 0} 
W_{D \pm}^\eps (q,p) \, T(a) \, W^{\eps \,*}_{D \pm} (p)   
= \Omega_\pm \, \alpha_{as} (T(a)) \eqq \Omega_\pm \, (T_{as}(a)) \, . 
 \ \ \endproof   $$

\vspace{1mm}

Summarizing, for the above model, describing classical particles with Coulomb 
interaction and with realistic, translation invariant, coupling to the 
quantized electromagnetic field, the introduction of an asymptotic reference 
dynamics $U_D(t)$ a la Dollard, an adiabatic switching and a particle energy 
renormalization term, we obtain the existence of the  
M{\o}ller operators and of 
the scattering matrix, describing infinite photon emission.

The M{\o}ller operators interpolate between the dynamics, $U(t)$, and the
asymptotic dynamics, $U_{as} \eqq U_+(t) = U_-(t)$, 
uniquely associated to $U_D(t)$ by eq.\,(5.17);
the $S$-matrix is invariant under $U_{as}(t) $ and $T_{as}(a)$. 

The M{\o}ller operators 
and the $S$-matrix exhibit a factorization of
the infrared divergences which may also be displayed for the particle
scattering. In fact, the same M{\o}ller operators are obtained if
an adiabatic switching is adopted also for the particle Coulomb interaction,
as discussed in Appendix D.

\vspace{2mm}
We have therefore, 

\begin{Theorem}
{For the model defined by eqs.\,(2.1)-(2.6) and (5.6), with the adiabatic 
regularization given by eqs.\,(4.1),\,(4.2) and (Dollard) reference dynamics
$U^\eps_D(t)$ given by eqs.\,(3.5),\,(3.7),\,(4.9),\,(4.10), 
or with $h_D$ replaced by $h_D^\eps$, eq.(D.1), one has:
 
\noindent
i) the M{\o}ller operators exist as strong limits 
\be{
\Omega_\pm = 
\lim_{\eps \to 0} \lim_{t \to \pm \infty} \Omega^\eps_t \ , \ \ \ 
\Omega^\eps_t \eqq U^{\eps\,  *}(t) \, U_D^\eps(t),}\ee
on the whole Hilbert space (with $L^2(\dR^{6N})$ replaced by
$L^2(\Gamma_{nr})$, eq.\,(2.8), in the n.r. case);

\noindent
ii) the asymptotic dynamics $U_{as}(t)$ associated to the Dollard reference 
dynamics, eq.\,(5.17), is a one-parameter continuous group; 
it is the product of the free particle 
dynamics and a non-Fock coherent transformation of the free photon dynamics, 
eq.\,(5.22), indexed by the 
momenta of the particles;
the M{\o}ller operators interpolate between the dynamics $U(t)$ and 
the aymptotic dynamics
$U_{as}(t) $ 
\be{ H \, \Omega_\pm = \Omega_\pm \, H_{as} \ , \ \ \ \ 
H_{as} \eqq h_0 + \alpha_{as}(H_0) \, ; }\ee

\noindent
iii) the scattering matrix, 
$ S= \Omega_+^* \, \Omega_- $, commutes with the asymptotic dynamics 
$U_{as}$ and with the asymptotic space translations $T_{as}$. 

\noindent
For two particles with repulsive Coulomb interaction, $\Omega_\pm$ and $S$
are unitary operators; }

\noindent
iv) the M{\o}ller operators explicitly display the photon infrared 
divergences, 
\be{
\Omega_\pm  = \lim_{\eps \to 0}
W_{0 \pm}^\eps (q,p) \, W^{\eps \,*}_{D \pm} (p_{\pm}) \,\om_\pm 
=  \lim_{\eps \to 0}
W_{0 \pm}^\eps (q,p) \, \om_\pm  \, 
W^{\eps \,*}_{D \pm} (p) \, ; }
\ee
in fact, $ W_{0 \pm}^\eps (q,p) \,\omega_\pm$ 
are the M{\o}ller operators relative to the 
free photon dynamics, eq.\,(4.15),  and their infrared divergences 
are canceled by the (time-ordered) non-Fock coherent factors 
$W^{\eps \,*}_{D \pm} (p)$;

\noindent
v) the explicit factorization of all the infrared divergences is displayed
in the following form of the $S$-matrix:
\be{
S = \lim_{\eps \to 0}  W^\eps_{D +} (p) \, e^{ - i l^\eps \, \V_D} \, 
S_0^\eps \, e^{ - i l^\eps \, \V_D} \, W^{\eps \,*}_{D -} (p) \, , }\ee
with $S_0^\eps$ the standard adiabatic 
$S$-matrix, corresponding to the Hamiltonian
$$ H^\eps(t) = h_0 + e^{- \eps |t|} h_I + H_0 + H_{I, r}^\eps (t) \, , $$
\be{l^\eps \eqq \int_1^\infty  d s e^{- \eps s}/s \ , \ \ 
\V_D \eqq  \frac{e^2}{4 \pi} 
\sum_{i <j} \frac{v_i - v_j}{|v_i - v_j|^3} (V_i - V_J)\, . }\ee

\end{Theorem}


\section{LSZ asymptotic limits}

The model sheds light also on LSZ asymptotic limits in the
presence of Coulomb interactions and infinite photon emission.
For definiteness, we consider a system of identical charged (fermionic) 
particles, so that the Hilbert space is of the form
$ \H = \sum_n \H^n $, $\H^n = L^2_{ant}( \dR^{6n}) \oplus \H_F $,
$L^2_{ant} $ the space of $L^2$ functions of $n$ positions and momenta, 
antisymmetric under odd permutations.
The Hamiltonian is given on $\H^n $ by eqs.(2.4)-(2.6)
(the dynamics leaving invariant the antisymmetric wave-functions).

In the following, we adopt
the relativistic form of the velocity, the second of eqs.\,(2.3);
the same results hold in the non-relativistic case with suitable
domain and momentum space restrictions on the charged fields.


\subsection{Charged fields and their Dollard dynamics}

The charged fields $\Phi(q,p)$ are defined, on all $\psi^n \in \H^n$, 
$f \in \S(\dR^6)$, by
$$ (\Phi(f)^* \psi^n)(q, q_1 \ldots p \ldots p_n)  =  
\sqrt {(n+1)} \, ( f(q,p) \psi^n(q_1 \ldots  p_n) )_{ant} \, , $$  
the index $ant$ denoting the projection on the antisymmetric subspace,
and satisfy the anti-commutation relations
$$ \{ \Phi(q,p) , \, \Phi^*(q',p') \} = \delta (q-q') \, \delta(p-p') \  , 
\ \ \ \  \{ \Phi(q,p) , \, \Phi(q',p') \} = 0   \, .$$
In the following, it will be convenient to work with the partial
Fourier transform $\Psi^*(P,p)$ of $\Phi^*(q,p)$,
$$ \Psi^*(P,p) = 
(2 \pi)^{-3/2}\int d\,q \, e^{iqP} \, \Phi^*(q,p) $$
and, correspondingly, use states defined by antisymmetric 
$L^2$ functions $\psi(P_1,p_1 \ldots P_n, p_n)$.
As usual, 
$$ \Psi^*(f) \eqq \int \, dP \, dp \, \Psi^*(P,p) f(P,p),  
\ \ \ f \in \S(\dR^6)  \, ,$$
$$  \rho(P,p) \eqq \Psi^*(P,p) \Psi (P,p)\ , \ \ \ 
   \rho(p) \eqq \int d\,P \, \rho(P,p)  \, . $$ 
Then, 
$$ \rho(g) \eqq \int dP \,dp \, \rho(P,p) \, g(P,p)  \, , $$ 
and
$$
 \rho \rho (F) \eqq \int dr\, dr' \, \rho(r) \rho(r') F(r,r') \ ,
\ \ \ r \eqq (P,p) \, $$
are (unbounded) multiplication operators in Fock space, 
for all measurable $g, F$, with $F(r,r)$ measurable, and satisfy,
for $F(r,r') = F(r',r)$,
$$ 
e^{i \rho(f)} \, \Psi^*(P,p) \, e^{-i \rho (f)} = e^{if(P,\,p)} \, 
\Psi^*(P,p) \, , $$
$$ 
e^{\frac{i}{2}
 \, \rho \rho (F)} \Psi^*(P,p) \, e^{-\frac{i}{2} \,  \rho \rho (F)} = $$
$$ = \Psi^*(P,p) \, 
e^{i \int dP' \, dp'\,\rho(P', \, p') F(P,\,p,\,P', \,p')  + \frac{i}{2} \, 
F(P,\,p,\,P,\,p) } \, , $$

The use of the Wick ordered product $: \rho \rho : $ leads to the same equation
with the omission of  $e^{\frac{i}{2} \, F(P,\,p,\,P,\,p) } $.

The free particle Hamiltonian takes the form
$$ h_0 = \int dP \, dp \, \Psi^*(P,p)  P v(p) \Psi(P,p)  $$
and 
\be { u_0 (t) \Psi^*(f) u_0^*(t) = \Psi^*(f_{-t}) \ , 
\ \ \ f_{-t} (P,p) = e^{i P v t}f(P,p) \, ; }
\ee
clearly, $\rho(P,p) $ is invariant under the free evolution.

The Dollard evolution operator
$U^\eps_D (t) = u_D(t) \,\U_0(t)\, \U^\eps_D(t)$, with $\U_0$ the free
electromagnetic evolution, is given by 
(see eqs.\,(3.4), (4.11)-(4.14) and the redefinition in Proposition 5.1),
\be { u_D(t) = u_0(t) e^{\frac{i}{2} \, : \rho \rho : ( C_t ) } \, ,} 
\ee
\be { C_t \eqq  \frac{ e^2}{  4 \pi }\, {\SI} \, t \, \ln|t| 
\frac{v-v'}{|v-v'|^3} \,(V-V') 
\eqq {\SI} \, t \, \ln|t| \, C(P,\,p,\,P',\,p')
\, , }
\ee
$$ \U^\eps_D(t) = e^{ - i \int dp \, [ a(F^{D \eps}_t(p)) + h.c. ] \rho(p) } \,
e^{ \frac{i}{2} \int  dp \, dp' \, L_t^\eps(p,p') \rho(p) \rho(p') }\times $$
$$ \times e^{i \int dp \, \rho(p) \, 
(\delta E(p) \, \int_0^t ds \, e^{-2 \eps |s|} + \delta \varphi^{D \, \eps}_\pm (p)) }
\, , $$ 
$$ F^{D \eps}_t(p) = F^{D \eps}_t(k, \lambda, p) \eqq
F^{D \eps}_\pm(k, \lambda, p) (1 - e^{- \eps |t|} e^{- i (|k| - vk) t}) \, ,$$
with 
$F^{D \eps}_\pm(k, \lambda, p)$ ($ \pm = \SI \, t $)  
the one-particle coherent factor given by eq.\,(4.17),  
\be { L^\eps_t(p,p') \equiv \ume 
 \, \int_0^t ds \, [ G^\eps(s,p,p') + G^\eps(s,p',p) ] \, , }
\ee
 $ G(s,p,p') $ given by  eq.\,(5.5),
$\delta E (p)$ by eq.\,(5.6). 

Since $u_D $ and $\U^\eps_D$ are
multiplication operators in the $P,p$ representation, 
they commute with $\rho(P,p)$ and therefore
$$ [U^\eps_D (t) , \, \rho(P,p) ] = 0 \, .
$$

The Dollard dynamics of the field $\Psi(f)$ is therefore given by
\be { \Psi^{\eps \, *}_D(f,t)  
\eqq   U^\eps_D (t)\, \Psi^*(f)\, U^{\eps \, *}_D (t) = }
\ee 
$$
 = \int dP \, dp \, f_{-t}(P,p) \, \Psi^*(P,p) \, 
e^{ - i  [a (F^{D \eps}_t(p),t) + h.c. ]  } \,
e^{i \rho \, (\chi^\eps_t(P,p))} \, , $$
where
$$
a (F^{D \eps}_t(p), t) 
\eqq \sum_\lambda \int dk \, a(k, \lambda) 
F^{D \eps}_t(k,\lambda,p) \, e^{i k t} \, ,
$$
\be{ \rho \, (\chi^\eps_t(P,p)) = 
\int dP' \, dp' \,\rho (P',p') ( L^\eps_t (p,p') + C_t(P,p,P',p')  
+ c_t^\eps (p, p')) \,  ,  }
\ee
$ c_t^\eps (p, p') \equiv {\rm Im} 
( F^{D \eps}_t (p), F^{D \eps}_t (p'))$,    
having a finite limit for $t \to \infty$ and then for $\eps \to 0$;
a ``diagonal phase'' has been omitted since it vanishes for 
$t \to \pm \infty$, see Proposition 5.1.


The integration in eq.\,(6.5) is well defined since the exponential
is strongly continuous in $P,p$ by Lemma 4.2 and eqs.\,(6.2), (5.5)
(on a dense domain and therefore everywhere).
$\rho \, (\chi^\eps_t) $ describes the (logarithmically divergent) 
Coulomb phases, with $ C_t $ arising from the (classical) Coulomb 
interactions of the particles and
$L^\eps_t $ representing the Lienard-Wiechert corrections produced
by the interaction with the photons. 

For the electromagnetic field we have
\be { 
a^{\eps \,*}_D (k, \lambda, t) \eqq 
U^{D \eps} (t) \, a^*(k, \lambda) \, U^{D \eps \, *} (t) = } \ee
$$ = e^{- i |k| t} \, a^*(k, \lambda) 
- i \int dp  \,F^{D \eps}_t (k,\lambda ,p) \, \rho(p) \,  , $$
on the sum over $N$ of the $N$ particle domains $D$ introduced before
($C^1$ wave functions of compact support with values in $D(H_0)$,
see Prop. 4.1), still denoted by $D$. $D$ is also stable under $U_{as}(t)$, 
as a consequence of Lemma 4.2.

\subsection{LSZ asymptotic limits of Heisenberg fields}

\noindent
\emph{a. Charged fields}
\vspace{1mm}

\noindent
The Heisenberg asymptotic charged fields $\Psi_{out/in} (P,p)$
are defined by
$$  \Psi^*_{out/in} (f) = \Omega_\pm \, \Psi^*(f) \, \Omega^*_\pm   =
\lim_{\eps \to 0} 
\Omega^\eps_\pm \, \Psi^*(f) \, \Omega^{\eps *}_\pm  $$
on $ \H_\pm 
= \Omega_\pm \, \H $,
see eq.\,(5.8).
By eq.(6.5),
\be { \Psi^*_{out/in} (f) = \lim_{\eps \to 0} \lim_{t \to \pm \infty}
U^{\eps \, *} (t) \, \Psi^{\eps \, *}_D(f,t)  \, U^\eps (t) \, . }
\ee
Thus, for the construction of the Heisenberg asymptotic field, the r\^ole
of the Dollard dynamics is to provide explicit corrections to the free
evolution, which allow for the existence of the asymptotic limits;
the main virtue of the Dollard correction is to subtract the infrared divergent
terms which arise in the standard formulation.

It should be stressed that, while in the standard case the interaction picture
free fields are isomorphic to the asymptotic Heisenberg fields, the 
``interaction picture'' Dollard fields 
$ \Psi^{\eps}_D(f,t) $, $a^{\eps}_D (k, \lambda, t)$,  which strictly    
correspond to the fields introduced by Rohrlich for QED  \cite{JR},
have little to do with the asymptotic fields; 
in particular, their time evolution is substantially different, see below.

Eq.(6.8) can be written in an (``adiabatic'') LSZ form. To this purpose, 
we note that $ a (F^{D \eps}_t(p), t) + h.c. $ can be written in terms of
the usual invariant smearing (in the space variables) 
of the electromagnetic potential $A $ 
with the Green function $D \, \theta (x_0)$ of the wave equation, 
\be 
{ a (F^{D \eps}_t(p), t) + h.c. =  e \, v(p) \int_0^t ds \,e^{-\eps |s|} \,  
A \, (\stackrel{\leftrightarrow}{\partial_t} D_{t-s} * \tilde \eta_{v(p)s}) \, , }
\ee
where 
$ A \, (\stackrel{\leftrightarrow}{\partial_t} D) \eqq  
- A (\dot D) + \dot A (D) $ and $ \tilde \eta_{v(p) s} (x) 
\eqq  \tilde \eta (x - v(p) s) $. 

The r.h.s. of eq.\,(6.9) has a simple physical interpretation since
$ Y^\eps_\mu(x,t;v) \eqq  \int_0^t ds\,e\,v_\mu \,e^{-\eps |s|}\, \,  
 (D_{t-s} * \tilde \eta_{v s})(x) $ is the Lienard-Wiechert potential generated
at time $t$, with vanishing Cauchy data at $t =0$, by 
the current $ j_\mu^\eps(v; x, s) \eqq e \, v_\mu \tilde \eta(x-vs) \, 
 e^{- \eps |s|} $,  $v_\mu \eqq (1, v)$. 

Then, denoting by $ \Psi^{\eps}_t$, $A^{\eps}_t $, $\rho^\eps_t $ 
the Heisenberg time evolution of $\Psi$, $A $ and $\rho$, 
under $U^\eps(t)$,
we have, on $\H_\pm$,\goodbreak
$$ \Psi^*_{out/in} (f) = \lim_{\eps \to 0} \lim_{t \to \pm \infty}
  \int dP \, dp \, f_{-t}(P,p) \, \Psi^{\eps \, *}_t(P,p) \, 
 e^{i \rho^\eps_t \, (\chi^\eps_t(P,\,p))} 
$$
\be { \times  \exp {-i  \int_0^t ds\,   
 A^{\eps}_t ( \stackrel{\leftrightarrow}{\partial_t} 
D_{t-s} * j^\eps(v(p); s))} 
\, . }
\ee

Eq.\,(6.10) provides an explicit modification of the standard
LSZ prescription for the asymptotic limit of the charged fields.
It amounts to the insertion of Coulomb phases and 
of the exponential of an electromagnetic operator,
both given by fields at time $t $, smeared with explicitly given 
test functions.

The effect of the electromagnetic factor is to
provide a shift of the electromagnetic potential at time $t$ by  
the Lienard-Wiechert potential $ Y^\eps(x,t;v) $ produced in a Huyghens
cone by the above current $ j_\mu^\eps $.
The Coulomb phases and the exponential of $A$ commute, since so do
the corresponding terms in eq.\,(6.5).

We stress that the main achievement of the LSZ asymptotic limit,
with respect to the interaction picture approach, is fully reproduced 
in eq.\,(6.10), which has a well defined meaning (within the adiabatic 
approach) independently of the existence of the M{\o}ller operators.

The strong convergence
of the density operators on $ \Omega_\pm \H^n $, $\forall n$,
for $t \to \infty$ and then $\eps \to 0$, 
$$  \rho^\eps_t(F) 
\eqq U^{\eps \, *} (t) \,  \rho(F) \, U^{\eps } (t) \to  
\Omega_\pm \, \rho(F) \, \Omega_\pm^* \eqq \rho_{out/in}(F) \, ,$$  $F(P,p)$ 
bounded, 
follows from the invariance of $\rho(P,p) $ under the Dollard evolution
and the norm boundedness of $\rho(F)$ on $\H^n$.
Clearly, $\rho_{out/in}(P,p) = \Psi^*_{out/in}(P,p) \Psi_{out/in}(P,p)$;
both $\rho_{out}(F)$ and $\rho_{in}(F)$ define commutative algebras. 

Furthermore, by the same argument, if $F^\eps_t$ converges 
uniformly to $F_\infty$,
then 
\be { \rho^\eps_t(F^\eps_t) \to  \rho_{out/in}(F_\infty) \ \ , \ \ \ 
 e^{i \rho^\eps_t(F^\eps_t)}  \to  e^{i\rho_{out/in}(F_\infty)} \, ,}
\ee
strongly on $ \Omega_\pm \H^n $, $\forall n$;
by a density argument, the second of eqs.\,(6.11) only requires
the uniform convergence of $F^\eps_t$ on compact sets.
\goodbreak


\vspace{2mm}

\noindent
\emph{b. Electromagnetic fields}
\vspace{1mm}

\noindent
For the asymptotic limit of the electromagnetic field,
it is convenient to work with their Weyl exponentials,
$$ 
W(f, \lambda) 
\eqq e^{ -i (a (f,\, \lambda) + h.c. )} \, .
$$  
Then, for (complex) $f$ and $|k|^{-1/2} f$
in $L^2$, on $\H_\pm$,
$$
W_{out/in} (f, \lambda) \eqq 
\Omega_\pm \, W(f, \lambda) \, \Omega^*_\pm   
\eqq e^{ -i (a_{out/in} (f,\, \lambda) + h.c. )} =
$$
\be {
\lim_{\eps \to 0} \lim_{t \to \pm \infty} 
U^{\eps \, *} (t) \, U_D^\eps (t) \, 
W(f, \lambda) \, U_D^{\eps \,*}(t) \, U^{\eps} (t) \eqq 
\lim_{\eps , \, t} W^\eps_t (f, \lambda) \, . }
\ee
By eq.\,(6.7), 
$$
W^\eps_t (f, \lambda) = U^{\eps \, *} (t) \, W (f_{-t} , \lambda) \, 
e^{ -i [ \int dp \, dk  \, \, i \, \overline{F^{D  \eps}_t} (k, \, \lambda, \, p) \, \rho (p)
\, f(k) + h.c.] } \, U^\eps (t) \, ,
$$
with $f_t(k) \eqq f(k) \, e^{- i \, |k| \, t}$.
As in eq.\,(6.11), the second factor converges 
for $t \to \pm \infty$ and $\eps \to 0$, to 
$$ 
\exp \, ( {i \rho_{out/in}  (J)(f, \lambda) + h.c. ))} \, ,
$$
where $J(k,\lambda,p)$, given by eq.\,(4.17) with $\eps = 0$,
is integrated with $\rho_{out/in} (p)$ and $f(k)$.
In fact, $k^{1/2} J \in L^2(d^3k)$, with norm bounded 
uniformly in $p$, so that, after integration in $k$, 
$ F^{D \, \eps}_\pm (p)$ converges uniformly for bounded $p$.

\noindent
Hence, also the first factor converges. Its limit 
is a unitary operator which is strongly continuous
in $\alpha$ for $f \to \alpha f$, since so are 
$W_{out/in} (\alpha f, \lambda) $ (by definition) 
and the limit of the second factor, by the above estimate.
Therefore,
\be { 
U^{\eps \, *} (t) \, W (f_{-t}) \, U^\eps (t)  \to 
e^{-i (b_{out/in}(f, \, \lambda) + h.c.)} \, . }
\ee
Since $ a_{out/in}$, $ a^*_{out/in}$, briefly  $ a^\#_{out/in}$ and  
$\rho_{out/in}  ( J)(f, \lambda) $ are well defined on 
$\Omega_\pm D$, on such a domain one has
\be {  a^*_{out/in} (f, \lambda) = b^*_{out/in} (f,\lambda) 
 - \rho_{out/in} (J)(f, \lambda) \, . }
\ee
Eq.\,(6.14) states that $b_{out/in}^\#$ are related to $a^\#_{out/in}$
by the second quantized version of the transformation of eq.\,(5.18).

\noindent
$b^\#_{out/in} (f,\lambda)$ define {\em free massless fields}; 
in fact, by eq.\,(5.19), for $t \to \infty$ and then $\eps \to 0$, 
$$ U^*(\tau) \, U^{\eps \, *} (t) \, U^{\eps } (t+\tau) 
\eqq T^\eps (t+\tau, \tau) \to 1 $$
strongly and the same holds for its adjoint; therefore
$$ U^*(\tau) \, b^\#_{out/in} (f,\lambda) \, U (\tau) 
= \lim_{\eps , \, t} U^*(\tau) \, U^{\eps \, *}(t) \, 
a^\# (f_{-t},\lambda) \, U^{\eps}(t) \, U (\tau) = $$
$$ = \lim_{\eps , \, s}  T^\eps (s, \tau)  U^{\eps \, *}(s) \, 
a^\# (f_{\tau - s},\lambda) \, U^{\eps}(s) T^{\eps \, *} (s, \tau) =  
b^\#_{out/in} (f_\tau,\lambda) \, .  $$
We shall denote by $B_{out/in}(x,t)$ the corresponding fields in 
Minkowski space. They are the result of LSZ formulas 
for massless asymptotic fields, on the whole scattering spaces,
with no need of Dollard corrections,
in agreement with the general analysis by \cite{Bu} 
on the asymptotic limit in the massless case.


\vspace{3mm}

\noindent
\emph{c. Asymptotic algebras and space-time translations}
\vspace{1mm}

\noindent
In the following, when no confusion arises, the indexes $out/in$ shall be
replaced by the single index $as$; we only recall that $J$ and
$H_{as}$ are independent of the two alternatives. Since
$$
  [\rho_{as} (F), \,  \rho_{as} (G)] = 0 =
 [\rho_{as} (F), \,  a^\#_{as} (f, \lambda)]  \, , 
$$
both $B_{out}$ and $B_{in}$ satisfy the CCR and commute with 
$\rho_{out/in} (F)$, respectively.
As asymptotic field algebras 
we take  the polynomial algebra $\F_{as}$  generated
by the free photon fields $B_{as} (x,0)$, their time derivative 
and by the asymptotic charged fields
$\Psi^\#_{as}$, smeared, e.g., with test functions in $\S(\dR^3)$ and 
$\S(\dR^6)$, respectively.

By construction, eqs.\,(6.8),\,(6.12), $a^\#_{as}$ 
and $\Psi^\#_{as}$
are canonical independent fields (at equal times);
$a^\#_{as}$ is well defined on $\Omega_\pm D$, which is stable under
$\Psi^\#_{as}(f)$, for
$$\hat f(q,p)= (2 \pi)^{-3/2} \int dP \, e^{-iqP} f(P,p)$$
of compact support. Eq.\,(6.14) implies therefore 
the following commutation relations between $
B_{as}$ and $\Psi_{as}$:
\be {
[ b^*_{as}(k, \lambda), \,  \Psi^*_{as} (P,p) ] = 
  J (k, \lambda, p) \, \Psi^*_{as} (P,p)  \, ,}
\ee
\be {
[ b_{as}(k, \lambda), \,  \Psi^*_{as} (P,p) ] = 
   J (k, \lambda, p) \, \Psi^*_{as} (P,p)  \, .}
\ee
They hold, together with the equations for their h.c.,
on $\Omega_\pm D$, for the operators obtained by smearing
$ b^\#_{as}$ 
with $f$, $ k^{-1/2} f(k), k^{1/2} f(k) \in L^2$, and 
$\Psi^*_{as} $ with $g$, $\hat g$ of compact support.

They extend, by closure of the corresponding operators,     
to $g \in \S(\dR^6)$, since 
$ || k^{1/2} J(k, \lambda, p) ||_{L^2(d^3k)}$, 
is of order $|p|$. 
In particular, 
the commutation relations, eqs.\,(6.15), 
(6.16), hold for the fields which generate $\F_{as}$, all
smeared with test functions in $\S$, on
$D_{as} \eqq \F_{as} \psi_0$, $\psi_0$ the vacuum vector;
$D_{out/in}$ are dense in $\H_\pm$ by eq.\,(6.14) and 
cyclicity of the vacuum for the fields at $t=0$.

With respect to the standard case, the above 
non-standard commutation relations are the only modification, 
produced in the asymptotic 
algebras by the LSZ asymptotic formula for the charged 
fields, eq.\,(6.10).

By Proposition 5.3, the Hamiltonian is given, on 
$\Omega_\pm D(h_0 + H_0)$, by
$$
H = \Omega_\pm H_{as} \Omega^*_\pm = 
H_{as} (\Psi_{as}, \, a_{as}) = h_0(\Psi_{as}) + H_0 (b_{as}) = 
$$
\be {
\int dP \, dp \, v(p) \, P \, \Psi^*_{as} (P,p) \Psi_{as} (P,p) +
\sum_\lambda 
\int d^3k \, |k| \, b^*_{as}(k, \lambda) \,  
b_{as}(k, \lambda)  \, . }
\ee
Similarly, from eq.\,(5.21) one has, for the generator
of space translations, eq.\,(2.7),
\be { 
\P = \int dP \, dp \, P \, \Psi^*_{as} (P,p) \Psi_{as} (P,p)  +
\sum_\lambda \int d^3k \, k \, b^*_{as}(k, \lambda) \,   
b_{as}(k, \lambda) }
\ee
$$ \equiv P_c (\Psi_{as}) + P_{ph} (b_{as}) $$
In the above decompositions, eqs.\,(6.17),(6.18), the two terms commute,
due to eqs.\,(6.15),(6.16).
The commutativity of the two terms is also implied by the fact that 
$H_0 (b_{as})$ and $P_{ph} (b_{as}) $
implement the space-time translations of the free massless field 
$b_{as}$; this reproduces the structure advocated in \cite{FMS} 
in terms of explicit functions of the asymptotic fields,
$ H_{charge} = h_0(\Psi_{as}) $, $P_{charge} = P_c (\Psi_{as})$. 

The space-time evolution of 
$\Psi^*_{as} (P,p)$ follows from eqs.\,(6.15)-(6.18):
$$  U(a,t)^* \, \Psi^*_{as} (P,p) \, U(a,t) = 
$$
\be {
e^{ - b^*_{as} (J^{a,t}(p) - J(p))} \, 
e^{i P v(p) t} \, e^{-iPa} \,
\Psi^*_{as} (P,p) 
\, e^{ b_{as} (\bar J^{a,t}(p) -  J(p))} \, , }
\ee
with $J^{a,t} (k,\lambda,p) \eqq e^{i (kt - ka)} J (k,\lambda,p)$.
Hence, even if the Hamiltonian is the sum of two free Hamiltonians,
\emph{the time evolution of $\Psi^*_{as}$ is not free as a consequence of the
commutation relations, eqs.\,(6.15),\,(6.16)}.
Thus, on the $N$ charged particle states $\Psi_N$
obtained by applying
$\Psi^*_{as} $ to the vacuum,
$$  H \, \Psi_N \neq  h_0(\Psi_{as}) \, \Psi_N \, . 
$$


Given eqs.\,(6.17),\,(6.18), the non-comutativity of 
$ \Psi_{as}$ and $b_{as}$, eqs. (6.15), (6.16), 
is crucial for the absence of an eigenvalue at the bottom of
the spectrum of the Hamiltonian in the one-particle sector, 
at given $\P = P_{charge} + P_{ph} (b_{as}) $ and given 
particle momentum $p$.
In fact, in general $\P$ and $\sum_i p_i$ commute with $H$;
by eqs.\,(6.17),\,(6.18), on (improper) one-particle states $\psi$, 
with fixed $\P,p$,
$$ H \psi = (h_0(\Psi_{as}) + H_0 (b_{as})) \psi  =
 (\P \, v(p) + [H_0 (b_{as}) - P_{ph}(b_{as}) \, v(p)]  ) \psi  \, .$$
The operator in square brackets is positive since $|v(p)| < 1$
and the bottom of the spectrum of $H$ is an eigenvalue iff
$\psi$ is the vacuum vector for $b_{as}$; such a vector exists 
iff $J(k, \lambda, p) \in L^2(d^3k) $, 
which is excluded for almost all $p$.


It is also important to stress that neither the commutation relations nor the
time evolution of the asymptotic fields are affected by the Coulomb
and Lienard-Wiechert corrections in the LSZ procedure. \goodbreak

One may introduce the fields
\be {
\Psi^*_{as} (P,p,x)  \eqq
e^{-b^*_{as} (J^{x}(p) - J(p))} \, 
\Psi^*_{as} (P,p) 
\, e^{ b_{as} (\bar J^{x}(p) -  J(p))} \, , }
\ee 
which transform covariantly under space-time translations 
$U(a)$, $a = (a_i, a_0)$:
\be {
U(a)^* \, \Psi^*_{as} (P,p,x) \, U(a) =
e^{i P v(p) t} \, e^{-iPa} \,
\Psi^*_{as} (P,p, x+a) \, . }
\ee
Their commutation relations with 
$ b^\#_{as}$  are
\be {
[ b^\#_{as}(k, \lambda), \,  \Psi^*_{as} (P,p,x) ] = 
     e^{\pm i k x} \, J (k, \lambda, p) \, 
\Psi^*_{as} (P,p,x)  \, .}
\ee
The field algebra $\F_{as}$ identifies a unique $C^*$ algebra
$\A_{as}$, generated by $\Psi_{as}$ and the Weyl exponentials of
$b_{as}$,
$$ \W_{as} (f) = e^{-i ( b_{as}(f) + h.c.)} \, , $$
with $f = f(k, \lambda)$ a $C^\infty$ complex function of fast decrease.
Eqs.\,(6.15), (6.16) become
\be { \W_{out/in} (f) \, \Psi^*_{out/in} (P,p) \, \W_{out/in} (f)^* =
  \Psi^*_{out/in} (P,p) e^{2 i \, {\rm Re} J (f,\,p)} \, , }  
\ee
$$ J (f,p) =  \sum_\lambda \int d^3k \, 
J (k, \lambda, p) \, f(k, \lambda) $$

The algebras $\A_{out/in}$ have the structure of a semidirect product
of fermion and Weyl algebras, of the same form as that discussed by
Herdegen \cite{He1}\,\cite{He2}.
However, the time evolution is very different, since, in Herdegen algebra,
$\Psi_{as}$ is a free field (of definite mass). 
In fact, the time evolution of the Herdegen variables is given by  
the sum of the free Hamiltonian for $\Psi_{as}$ and a free (e.m.) 
Hamiltonian commuting with $\Psi_{as}$; in our algebra, this would be obtained 
by replacing, in eq.\,(6.17), $H_0 (b_{as})$ with $H_0 (a_{as})$. 
Moreover, the representation of the semidirect product algebra  
adopted by Herdegen differs from ours by the absence of the vacuum state,
which would give rise, in his case, to charged states of definite mass.


\vspace{2mm}

\noindent
\emph{d. Asymptotic form of the corrections to the LSZ formula}
\vspace{1mm}

The modification of the standard LSZ formulas for the charged fields
arising from the electromagnetic interaction can be written in terms 
of asymptotic e.m. fields and asymptotic currents.

\begin{Proposition} The asymptotic charged fields, eq.\,(6.10),
are also given by the following LSZ formula
$$ \Psi^*_{out/in} (f) = \lim_{\eps \to 0} \lim_{t \to \pm \infty}
  \int dP \, dp \, f_{-t}(P,p) \, \Psi^{\eps \, *}_t(P,p) \, 
 e^{i \rho^\eps_t \, (C_t(P,\,p))} \, 
$$
\be {  
e^{i \rho_{out/in} \, (L_\pm^\eps(p) + c(p))} 
\exp {-i B_{out/in} (j_\pm^\eps (v(p))) }
\, , }
\ee
on $\H_\pm$, with the notation of eq.(6.6), $C_t$ and $L^\eps_t$ given
by eqs.\,(6.3), (6.4), 
$ j^\eps_{\pm \, j} (v;x) \eqq j_j^\eps(v; x) \, \theta(\pm x^0) $, 
so that 
$$  B_{out/in} (j_\pm^\eps (v)) = e
\int_0^{\pm \infty}  ds \, e^{-\eps |s|} \, \int d^3x \, 
B_{out/in} ( x+  v s,s) \, v \, \tilde{\eta}( x)  \, ;$$
$$ c(p,p') \eqq \lim_{\eps \to 0} \int d^4x \, d^4y \, D_{ij} (x-y)
( - j_{\pm \, i} (v',x)  + 1/2 \,  j_{\pm \, i}^\eps (v',x) ) 
\, j_{\pm \, j}^\eps (v,y) \, ,
$$
with  $j_{\pm \, j} \eqq j_{\pm \, j}^{\eps =0} $, 
$\ v' \eqq v(p')$. 

\end{Proposition} 

\noindent \Pf \,
The basic content of eq.\,(6.24) is that, in eq.\,(6.10), 
one may first take the asymptotic limits of the fields 
$\rho^\eps_t$ and $A^\eps_t$, 
keeping $\eps$ and $t$ fixed in the test functions. 
The resulting procedure for the limits will be shown to give
the same result as the diagonal procedure, eq.\,(6.10),
apart from the correction given by $c(p)$. 

\noindent
1. We first control the $\eps, t$ 
limits of the fields in 
the phase and in the electromagnetic factor, keeping 
$\eps, t$ fixed in the smearing functions. 

\noindent 
Eqs.\,(6.13),\,(6.14) imply that
$$ 
V_A (p, \eps', \tau, \eps, t) \eqq e^{-i \int_0^t dx_0 \, 
A^{\eps'}_\tau (\stackrel{\leftrightarrow}{\partial_\tau} D_{\tau-x_0} * 
j_\pm^\eps (v(p),x_0) } 
$$ 
converges strongly to
\be { 
 e^{-i \int_0^t  dx_0
B_{out/in} (j_\pm^\eps (v(p),x))} =
 e^{-i (a_{out/in} (F^{D \,\eps}_t (p))  + h.c.)} \,
e^{- ( \rho_{out/in} ( i \, J ) ( F^{D \,\eps}_t (p)) \, -  \, h.c. ) } \,  } 
\ee
as $\tau \to \pm \infty$ and then $\eps' \to 0$, since
$ \int_0^t ds \, e^{- \eps |s|} \, D_{\tau - s} * \tilde \eta_{vs} $
is a regular solution of the wave equation, corresponding 
to $f_{-\tau}$ in eq.\,(6.13).

\noindent 
Similarly, $V_\rho (p, \eps', \tau, \eps, t) 
\eqq \exp {i \rho^{\eps'}_\tau \, (L_t^\eps (p) )} $
converges strongly, 
as $\tau \to \pm \infty$ and then $\eps' \to 0$, $\eps, t$ fixed, to
$$ V_\rho (p, 0, \pm \infty, \eps, t) 
\eqq \exp {i \rho_{out/in} \, (L_t^\eps (p) )} \, .
$$

\noindent 
2. 
We must prove that, 
in eq.\,(6.10), the phase and electromagnetic factor can
be replaced by their asymptotic version, apart from
a phase, i.e.,
on $\H_\pm$, omitting the $p$ dependence and using
$ [ \rho_{out/in}, \, b_{out/in}] = 0$,
\be {
V_\rho (\eps, t, \eps, t)   
V_A (\eps, t, \eps, t) - 
e^{i \rho_{out/in} \, (L_t^\eps  )}   
V_A (0, \pm \infty, \eps, t) 
\, e^{+i \rho_{out/in} (c - c^\eps_t)} \to 0 }
\ee
strongly  as $t \to \pm \infty$ and then $\eps \to 0$.

\noindent
Using
$ [ U^\eps_D (\tau) \, ,  \, \rho(p) ] = 0 $ and (see eq.\,(6.7))
$$
U^{\eps' \, *}_D (\tau) \, a(k,\lambda) \, U^{\eps'}_D (\tau)  = 
e^{-i k \tau} \, ( a(k,\lambda) - \rho (i \overline{F^{D \eps'}_\tau})  )(k,\lambda)) 
\, , $$
the first term in eq.\,(6.26) 
can be written as
\be { 
\Omega^{\eps}_t \, 
e^{i \rho (L^\eps_t (p))} \,
e^{-i (a (F^{D \,\eps}_t (p))  + h.c.)} \, \Omega^{\eps \, *}_t \,
e^{-  [ \rho_t  ( \overline{F^{D \eps}_t} ) ( F^{D \,\eps}_t (p)) \, -  \, h.c. ] } \,  . }
\ee

\noindent 
2a. The last factor in eq.\,(6.27) involves
the smearing of $\rho(p')$ with 
$-2i$ times 
the imaginary part of the scalar product
$ ( F^{D \,\eps}_t ( p'),\,  F^{D \,\eps}_t ( p)) $ as 
functions of $k$ and $\lambda$. Similarly for the 
last factor in eq.\,(6.25). 
By an explicit 
control, 
Im\,$( F^{D \,\eps}_t (p'), \,  F^{D \,\eps}_t (p))  $ 
converges, for $t \to \pm \infty$ and then $\eps \to 0$, to
$$ \lim_{\eps \to 0} \lim_{t \to \pm \infty} 
 \mbox{Im} \,(  -i J (p'), \,  F^{D \,\eps}_t (p))  
 - \ume  c^\eps_t(p, p') + \ume c(p,p')  \, ,$$ 
uniformly for $p,p'$ bounded.
Both terms are non-vanishing due to the presence of $\eps$ 
singularities in the corresponding integrals. 

\noindent
Then, by the second of eqs.\,(6.11), the last factor in
eqs.\,(6.27),\,(6.25) converge, for $t\to \pm \infty$ and then $\eps \to 0$,
and their limits differ 
by the factor $ e^{i \rho_{out/in} \, ( c'(p))} \,$, 
$\, c' = c - \lim_{\eps , t} c^\eps_t$. 
Both limits leave $\H_\pm$ invariant.


\noindent 
2b. We have to discuss the convergence of the remaining factors in
eq.\,(6.27), on $\H_\pm$. Since $\Omega^{\eps \, *}_t $ converges on $\H_\pm$, 
its limit inverts $\Omega_\pm$; eq.\,(6.26) 
reduces therefore to
\be {
(\Omega^{\eps}_t - \Omega_\pm) \, e^{i \rho (L^\eps_t)} \,
e^{-i (a (F^{D \,\eps}_t (p))  + h.c.)} \,  \to 0 \,  , \,\,\,
\mbox{strongly on} \, \H .}
\ee
Since, for $t \to \pm \infty$, $L^\eps_t (p')$ converges to $L^\eps_\pm (p')$ 
uniformly for bounded $p'$, $F^{D \,\eps}_t (k,\lambda, p) \to
F^{D \,\eps}_\pm (k,\lambda, p)$ in $L^2(d^3k)$ and
$\Omega^{\eps }_t  \to \Omega^{\eps }_\pm $, we are left with 
the limit in $\eps$.
Since $\Omega^{\eps}_\pm = W^{\eps}_\pm \, \omega_\pm \, $,
$ \, [W^{\eps}_\pm , \, \rho(p')] = 0 = [W_\pm , \, \rho(p')] \, $,
$\, \omega_\pm \, \rho = \rho_{out/in} \, \omega_\pm$, the
exponential of $\rho$ can be moved to the left, becoming
$ e^{i \rho_{out/in} (L^\eps_\pm)} $. 

\noindent
Since $ \omega_\pm  \, a (F^{D \,\eps}_\pm (p)) =
 a (F^{D \,\eps}_\pm (p_\pm)) \, \omega_\pm  $
and $\omega_\pm L^2(\Gamma) \times \H_F  = \H_\pm$, we are reduced to    
\be {
(W^{\eps}_\pm - W_\pm) \, 
e^{-i (a (F^{D \,\eps}_\pm (p_\pm))  + h.c.)} \,  \to 0  }
\ee
strongly on $\H_\pm$.
Using eq.\,(5.1) and the fact that, for any $n$-particle
subspace $L^2(\Gamma^{(n)}_\pm)$,
$W^{\eps}_\pm $ are multiplication operators $W^{\eps}_\pm (q',p') $, 
one has
$$ W^{\eps}_\pm \, 
e^{-i (a (F^{D \,\eps}_\pm (p_\pm))  + h.c.)} =
e^{- < \Delta F^{\eps}_\pm (q',p') , \, F^{D \,\eps}_\pm (p_\pm) >} \,
e^{-i (a (F^{D \,\eps}_\pm (p_\pm))  + h.c.)} \, W^{\eps}_\pm \, . $$
By the proof of Proposition 5.1, $|k|^{1/2} F^{D \,\eps}_\pm(p_\pm)$ 
and $|k|^{-1/2} \Delta F^{\eps}_\pm (q',p')$ 
converge in $L^2(d^3k)$, for almost all $(q',p')$ in $\Gamma_\pm$;
therefore, the above phases converge to
$$ \lim_{\eps \to 0}  \lim_{\eps' \to 0}   
< \Delta F^{\eps'}_\pm (q',p') , \, F^{D \,\eps}_\pm (p_\pm) > \, , $$
a.e. in $\Gamma_\pm$.
Eq.\,(6.27) then follows from 
$$ (W^{\eps}_\pm - W_\pm) \to 0 \ \ {\rm on} \ \H_\pm \ , \ \ \
  W_\pm \H_\pm = \H_\pm \, . \ \ \ \ \ \endproof 
$$

In eq.\,(6.24), the electromagnetic correction required for
the LSZ limit of the charged field is replaced by
a string-like factor involving the asymptotic photon
field $B_{out/in}$; the string is a straight line, with direction
given by the momentum variable $p$ of the test function $f$.

From this point of view, a convenient strategy for the asymptotic 
limit of the charged fields is to first obtain the massless asymptotic
photon field and then use the corresponding string as the e.m. correction
in the LSZ formula for the charged field.

In eq.\,(6.24), also the Lienard-Wiechert 
modification involves the asymptotic density $\rho_{out/in}$ 
of the particle momentum
$p$, whose construction does not require the LSZ limit of the charged field
(see eq.\,(6.11)). Since $L^\eps_\pm(p, p') = L^\eps_\pm(p', p)$, the
Lienard-Wiechert factor has no effect on 
the canonical structure of the asymptotic charged fields.

On the contrary, the finite phase factor involving 
$\rho_{out/in} (c(p))$ (not necessary for the existence of 
the LSZ limit) has an 
antisymmetric part, which cancels the changes induced 
by the exponential of the asymptotic e.m. field on the
charged field anticommutation relations. 

In order to write the Coulomb correction
$\rho_{out/in} (C_t(P,p))$ 
in terms of asymptotic variables, one should
adiabatically switch also the Coulomb potential, since
$ \exp{i \rho_{as} (C_t)} - \exp{i \rho_t^\eps (C_t)} $ 
does not converge for $t \to \pm \infty$.
The problem does not arise in a framework
without adiabatic switching \cite{MSpr}.


\section{Conclusions}

The strategy advocated by Kulish-Faddeev \cite{KF}
and Rohrlich \cite{JR} for QED, supported by
the cancellation of the infrared divergences in 
the perturbation expansion, has been rigorously
controlled in a translationally invariant model reproducing 
basic infrared problems of QED.

Technically, this has been obtained by introducing an
adiabatic procedure and mass renormalization counterterms; both ingredients
are characteristic of the Feynman-Dyson approach, in its non-perturbative
version discussed by Hepp \cite{Hp}.

The field theory version of the model provides a strategy for the
control of the asymptotic limit of the Heisenberg charged fields,
through an explicit modification of the LSZ (HR) formulas.
The resulting asymptotic fields are very different from those 
advocated by Zwanziger \cite{Z}, Schweber \cite{Sc} and 
Rohrlich \cite{JR}. 

The modifications of the LSZ formula may be written in two 
alternative forms: one is given in terms of the photon 
field at the same time $t$ of the charged field,
the other, by a string-like factor involving the 
massless asymptotic photon field.
The formulas only involve Heisenberg fields and 
geometrical factors; they represent the transcription of the KFR
strategy into the LSZ (Haag-Ruelle) approach and are therefore 
good candidates for asymptotic formulas in QED.
The same asymptotic fields can be obtained 
without adiabatic switching, through LSZ formulas 
with modified Dollard electromagnetic corrections
in coordinate space (subtracting the effects of logarithmic 
distortions of the trajectories in photon emission) \cite{MSpr}.

For the one-particle sector, our first LSZ formula is close to 
that proved by Chen, Fr\"ohlich and Pizzo \cite{CFP}, 
eqs.\,(III.29),\,(III.30), 
for one-particle states in non-relativistic QED.
In fact, their LSZ modification factor 
$W_{k, \sigma_t} (v,t)$ 
acts on a (previously constructed) one-electron state
which requires an infrared dressing by a Weyl operator $W$ of the 
same form, at $t=0$, in order 
to mimic the action of an interpolating field.
Then, modulo different infrared regularizations and other technical points, 
their correction factor 
$W_{k, \sigma_t} (v,t) \, W^{-1}$ corresponds to our e.m. correction 
for the one-particle case.

For the asymptotic limit of the electromagnetic field, 
the ordinary LSZ (HR) limit (with no Dollard correction) applies, 
in accord with Buchholz result \cite{Bu}, 
and defines massless fields $B_{out/in}$. 
The canonical fields 
$\Psi_{out/in}$ and $B_{out/in}$ 
generate asymptotic algebras with a semidirect product structure,
their commutation relations, eq.\,(6.15)\,(6.16), 
being determined by the 
electromagnetic field corrections to the LSZ formula for the 
charged fields. Their time evolution is
generated by the sum of the free Hamiltonians of
$\Psi_{out/in}$ and $B_{out/in}$, eq.\,(6.17).

Such a decomposition, which also holds for the momentum, reproduces the
splitting $P^\mu = P^\mu_{charge} + P^\mu_{ph}$, advocated in 
\cite{Fr}\,\cite{FMS}. 
The resulting structure is substantially different from the Herdegen
proposal of a semidirect product of asymptotic algebras 
\cite{He1}\,\cite{He2}, which involves
a different Hamiltonian, giving rise to free charged asymptotic fields.

The absence of charged states of definite mass, which in Herdegen
analysis requires the absence of the vacuum, 
follows here from 
eq.\,(6.17)
and the non-trivial commutation relations between 
$\Psi_{as}$ and $B_{as}$.

The space-time transformations of $\Psi_{as}$, $B_{as}$
are also different from those of the asymptotic fields 
proposed by Zwanziger \cite{Z}, mainly because his char\-ged 
fields include (Coulomb-Lienard-Wiechert) phase operators
which spoil the group property of their time dependence.

\goodbreak


\newpage


\appendix

\begin{center}
 {\Large \bf {Appendix}}  
\end{center}

\section{Asymptotics of classical configurations}

\begin{Lemma}

Let $ \gamma_t$, $t \in \dR$, be invertible measure preserving 
transformations of $\Gamma = \dR^{6N}$, with the Lebesgue measure
$dx$, defining therefore unitary operators $\omega_t$,
$\omega_t \psi (x) \eqq \psi(\gamma_t x)$, 
in $L^2( \Gamma, dx)$. If 
$\omega_t$ converge strongly, for $t \to \pm \infty$, to 
$\omega_\pm$, and
$\psi(\gamma_t x)$ converge pointwise, $\forall \psi \in D$,
$ D \eqq \cup_n C^1(A_n) $, $A_n$ open bounded sets covering
$\Gamma$ apart  from a set of zero measure, then
there exist measurable subsets $\Gamma_\pm$ and measure preserving
transformations $\gamma_\pm : \Gamma_\pm \to \Gamma$ such that:
\be { \gamma_t x \to_{t \to \pm \infty} 
\gamma_\pm x \ \ \forall x \in \Gamma_\pm \, ,}
\ee
$ (\omega_\pm \psi) (x)$ vanishes (a.e.) in the complement 
of $\Gamma_\pm$ and
\be { (\omega_\pm \psi) (x) 
= \psi(\gamma_\pm x), 
 \ \ \forall \psi \in L^2(\Gamma, dx) \ , \ \ x \in \Gamma_\pm \, ,}
\ee
\be { \omega_\pm L^2(\Gamma, dx) = L^2(\Gamma_\pm, dx) \, . }
\ee
\end{Lemma}

\noindent \Pf \, 
Let $\Gamma_\pm$ be the complements of the sets
$$ \{ x : \psi(\gamma_t \, x ) \to_{t \to \pm \infty} 0 \, , \ 
\forall \psi \in D \} \, .$$
$\Gamma_\pm$ are measurable since $D$ is separable in the Sup norm.
Let $x \in \Gamma_+$; then there exists $\psi_x \in D$ such that
$  \gamma_t \, x \in$ 
supp$(\psi_x)$, $\forall t > t_x$.
$\forall \epsilon > 0$, the (compact) support of $\psi_x$ can be 
covered by a finite number of balls $B_i^\epsilon$, of radius $\epsilon$,
and a partition of unity argument shows that for some index $i$,
$  \gamma_t \, x \in B_i^\epsilon$ for all large $t$, so that $\gamma_t \, x $
has the Cauchy property; we denote by  
$ (\gamma_+ \, x)$ its limit.
For $\psi \in D$, eq.\,(A.2) follows by the identification of $L^2$ limits
with pointwise limits. 
Since $D$ is dense in $L^2$ and $\omega_+$
is an isometry, $\gamma_+$ preserves the measure, so that eq.\,(A.2)
extends to $L^2$ and the image of $\omega_+$ can be identified 
with  $L^2(\Gamma_\pm, dx) $. $\ \ \ \endproof $

\vspace{2mm}

\section{Completeness of the 
M{\o}ller operators in the repulsive two-particle case}

\noindent We consider the non-relativistic case,
the relativistic case being very similar.
For $N=2$, in the reference 
frame where $p_1 = - p_2 \eqq p$,
all trajectories have non-zero relative asymptotic velocity $v_\pm$
and therefore 
\be { |x_t| \eqq | q_{1\,t} - q_{2\, t} | \geq (1- \epsilon)\, |v_\pm| \, |t| 
\ \ \ \ \ \ \mbox{for}\ |t| \ \  \mbox{large} \, . } 
\ee 
This allows for the existence of the limit of
$u_D^*(t) \, u(t) \, \psi$ as $t \to \pm \infty$, 
which implies the unitarity of $\om_\pm$. 
In fact,
\be { (d/dt) \, (u_D^*(t) \, u(t)) \, \psi = u_D^*(t) \, u(t) \, 
(w(x_t) X(t) - w(v_t t ; 0) V(t) \, t )  \, \psi \, ,}
\ee 
where 
$$ v \eqq \dot q_1 - \dot q_2 \, , \,\,\,\,\,\ V \eqq V_1 - V_2 \, , 
\,\,\,\,\,X \eqq (Q_1 - Q_2) \, , $$
\be { X(t) = u^*(t) X u(t) \ , \,\,\,\, \ V(t) = u^*(t) V u(t) \, . }
\ee
Now, by eq.\,(3.14), $u(t)$ induces a linear transformation on $X, V$, 
with coefficients given by matrices $A(t) = A_{\alpha\, \b }(q,p,t)$, etc., 
$\alpha, \b = 1,2,3$,
\be {X(t) = A(t) X + B(t) V \ , \ \ \,\,\,
     V(t) = C(t) X + D(t) V \, , }\ee
which (in the non-relativistic case, with reduced mass $= 1$) satisfy
$$  \dot A = C \ , \ \ 
      \dot B = D \ , \ \  $$
$$  \dot C_{\alpha \b}(t) = - \V^{''}_{\alpha \gamma}(t) \,A_{\gamma\, \b} (t)\ , \ \ \   
\dot D_{\alpha\, \b}(t) = - \V^{''}_{\alpha \gamma}(t) \,B_{\gamma\, \b} (t)\ , 
$$
with $ \V^{''}_{\alpha \gamma} \equiv  
\partial^2 \V / \partial x_\alpha \partial x_\gamma$ and summation over repeated
indices. 
Then, 
$$ C(t) = C(t_0) - 
\int_{t_0}^t ds \, \V^{''}(s) \, A (t_0) - 
    \int_{t_0}^t ds  
\int_{t_0}^s ds' \,  \V^{''}(s) \, C(s') \,  . $$  

\noindent By eq.\,(B.1),  
$\V^{''}(t) = O(t^{-3})$, so that 
$$ \sup_{t \geq t_0} || C (t) || 
\leq || C (t_0 ) || +     || A (t_0 ) || \,  O(t_0^{-2}) +  
\sup_{t \geq t_0} || C (t) || \, O(t_0^{-1}) $$ 
and therefore
\be { \sup_{t \geq t_0} || C (t) || 
\leq || C (t_0 ) || \, (1 + O(t_0^{-1})) \, . }
\ee
This implies $|| A (t) || = O(t)  $ and 
$|| \dot C (t) || = O(t^{-2})  $. 

\noindent
Then, 
$ || A (t) - C (t) t || = O(\ln t)  $. The same conclusion holds for 
$ B(t)$ and for $ D(t)$. 
This yields the estimate, $\forall \psi \in D^1_0$ (see Proposition 3.1)
\be{ || (w(x_t) (X(t) -  V(t) \, t ) \, \psi ||
 = O(t^{-2} \ln t) \, .}
\ee
On the other hand,
\be{ || (w(x_t) - w(v_t t ; 0)) \, V(t) \, t ) \, \psi ||
 = O(t^{-2} \ln t) \, }
\ee
since $|| V(t) \psi ||$ is bounded by eq.\,(B.5) for $C(t)$
and $D $ and eq.\,(B.1) implies
$$ | x_t - v_t t  | = O (\ln t) $$
on the support of $\psi$, which is left invariant by $V(t)$.
Then, the argument following eq.\,(3.17) applies.

\vspace{2mm}

\section{Proof of Lemma 4.2}

Existence and unitarity  of $U(f)$, $\forall f \in \ltk$, 
follows from the essential 
self-adjointness of $a(f) + a(f)^*$ on the domain $D_{fin}$ of vectors 
describing 
finite numbers of particles. If $f, |k|^{-1/2} f \in \ltk$, 
then $a(f) + a(f)^*$ is well 
defined on $D(H_0)$ as a consequence of the following estimates 
for $a^\#$, $a^\# = a, 
a^*$, on $D_{fin}$ and therefore on $D(H_0)$:
$$ ||\adie(f) \Psi||^2 \leq ||k^{-1/2} f ||^2 (\Psi, \,H_0 \,\Psi) 
+ || f||^2  \leq $$
\be{\leq ||k^{-1/2} f ||^2 (a \,||H_0\Psi||^2 + 
(1/4a) ||\Psi||^2) + ||f||^2 ||\Psi||^2, \,\,\,\,a > 0.}\ee 
\goodbreak
\noindent Furthermore, under the above assumptions 
of differentiability, one has 
$$U(f_{\a + \epsilon}) - U(f_\a) = U(f_\a + \epsilon f'_\a + 
\epsilon g_\a(\epsilon)) - 
U(f_\a),$$
with  $g(\epsilon), \,|k|^{-1/2}\,g(\epsilon) \ra 0$ in 
$L^2(d^3 k)$ as $\epsilon \ra 0$, which can also be written as
$$ U(f_\a)\,[\, U(\epsilon f'_\a) U(\epsilon g_\a(\epsilon))\,
e^{ (\epsilon \,C_\a + \epsilon \,o(\epsilon) )} - \id].$$
\noindent 
Now, by eq.\,(C.1), $\forall \psi \in D(H_0)$, 
$h, |k|^{-1/2} h \in \ltk$
$$ || (d/d\l) \,U(\l h) \psi ||^2  \leq 2\,||h \kum||^2 \, 
( ||H_0 \psi||^2 + || \psi||^2) + 4 ||h||^2 \, || \psi||^2, $$ 
which implies
\be { || (U(h) - \id)\,\psi|| \leq || h \kum || \, 
c_\psi + ||h || \, d_\psi \, . }\ee
Then, for $h = \epsilon\,g_\a(\epsilon)$, $\forall \psi \in D(H_0)$,
$ \epsilon^{-1}\,( U(\epsilon g_\a(\epsilon)) - \id ) 
\psi \to 0$, as $
\epsilon \to 0.$ 
On the other hand, by Stone theorem, on $D(H_0)$
$$ \epsilon^{-1} \,(U(\epsilon f_\a') - \id ) \ra i \, (a(f_\a')  + a(f_\a')^*) 
$$
so that, on $D(H_0)$,
$$d U(f_\a)/d \a = U(f_\a) \,
[i(a(f_\a')  + a(f_\a')^*) + C_\a]$$ and eq.\,(4.20) follows. 

\noindent 
Finally, $f_t(k) = f(k) \, e^{- i |k| t}$ satisfies the 
above conditions for $f_\a(k)$, if $f(k), 
\kum f(k), |k| f(k) \in \ltk$ and therefore,  $\forall \psi \in D(H_0)$
\be{ e^{ i H_0 t}\, U(f)\,\psi = U(f_{-t})\,e^{i H_0 t}\,\psi }
\ee
is differentiable in $L^2$ with respect to $t$; this implies 
$U(f)\,\psi \in D(H_0)$.

\vspace{2mm}

\section{Adiabatic switching of the Coulomb interaction}
 
In order to display the complete factorization of the infrared divergences,
we introduce the Hamiltonian
\be{ 
\tilde H^\eps = h_0  + H_0 + e^{- \eps |t|} h_I + H_{I,\, r}^\eps (t) \, . }
\ee 
We proceed as in Sect.(3.4), with $H^{\eps}_D (t)$
replaced by
$$ 
H^{\eps, \eps'}_D (t) =  h_0  + H_0 +
e^{- \eps' |t|} h_I(vt, Vt; 0) + H_{I,\, D}^\eps (t)  \, , 
$$
with $ H_{I,\, D}^\eps (t) $ still given by eq.\,(4.2). 
For $t \to \pm \infty$, the result are
the M{\o}ller operators 
$$
\Omega^{\eps, \eps'}_\pm =
W_{0 \pm}^{\eps, \eps'} \,\om^{\eps'}_\pm \,W^{\eps \,*}_{D \pm} =
W_{0 \pm}^{\eps, \eps'} (q,p)\,W^{\eps \,*}_{D \pm} (p^{\eps'}_\pm) \;\om^{\eps'}_\pm 
$$
with 
$p^{\eps'}_\pm\, \om^{\eps'}_\pm = \om^{\eps'}_\pm  \,p $, as in
eq.\.(3.12).

\noindent
In fact, the estimate eq.\,(3.16) holds uniformly in $\eps'$
and therefore the limits in eqs.\,(3.9)-(3.11) are uniform in $\eps'$.
This also implies that 
$\om^{\eps'}_\pm $ converges to $\om_\pm $ as $\eps' \to 0$.
Moreover, with an obvious extension of the notation of Appendix A,
as a consequence of the adiabatic cutoff, both
$\gamma^{\eps'}_t (q,p) $ 
and its inverse converge pointwise a.e. in $\Gamma$ as $t \to \pm \infty$.
The limit of the first, $\gamma^{\eps'}_\pm (q,p) $, satisfies
$$
(\om^{\eps'}_\pm \psi) (q,p) = \psi (\gamma^{\eps'}_\pm (q,p)) \, .
$$
Convergence of $\gamma^{\eps'}_t (q,p)^{-1} $ a.e. in $\Gamma$ 
implies $\Gamma^{\eps'}_\pm = \Gamma$
and then eq.\,(A.3), applied to $\om^{\eps'}_\pm $, implies that
$\om^{\eps'}_\pm   $ are unitary operators.
By the above uniformity argument, 
eq.\,(A.1) applies to $\gamma^{\eps'}_\pm $, i.e., 
for $\eps' \to 0$,
$$ \gamma^{\eps'}_\pm (q,p)  \to \gamma_\pm (q,p)  $$
a.e. in $\Gamma_\pm$. In particular, 
$$
p^{\eps'}_\pm (q,p) \to p_\pm (q,p) \  , \ \ \ \ \forall (q,p) \in \Gamma_\pm
$$
and eqs.\,(3.18),\,(3.19) hold uniformly in $\eps'$.
This allows for the control of the convergence of the $W$ operators
as in Sect.(5.1), see  Proposition (5.1):
$$
\lim_{\eps \to 0} W_{0 \pm}^{\eps, \eps} (q,p)\,W^{\eps \,*}_{D \pm} (p^\eps_\pm) =
\lim_{\eps \to 0} \lim_{\eps' \to 0} 
W_{0 \pm}^{\eps, \eps'} (q,p)\,W^{\eps \,*}_{D \pm} (p^{\eps'}_\pm) = W_\pm \, .
$$
In fact, the estimates for the convergence of $\Delta F^{\eps, \eps'}_\pm$
in the proof of Proposition 5.1 only rely on the limits in 
eqs.\,(3.18),\,(3.19), which are uniform in $\eps'$. 

\noindent Such uniformity
also implies that the estimates of  eqs.\,(5.13),\,(5.14), and therefore
the convergence of the phases, are uniform in $\eps'$.
In conclusion,
$$ \lim_{\eps \to 0} \Omega^{\eps, \eps}_\pm = 
\lim_{\eps \to 0} \Omega^{\eps}_\pm = \Omega_\pm \, . $$
By an explicit calculation, the M{\o}ller operators 
$ \omega^\eps_\pm $ 
are related to the standard adiabatic
M{\o}ller operators $ \omega^\eps_{0 \, \pm} $,  defined solely by an
adiabatic switching of the Coulomb interaction
with no Dollard correction, by 
\be { 
\omega_\pm = \lim_{\eps \to 0} \,  
\omega_{0 \, \pm}^\eps \, e^{\pm i l^\eps \, \V_D} \, , 
\ \ \ l^\eps = \lim_{t \to \infty} l^\eps_t}
\ee
$$ l^\eps_t \eqq \int_1^t e^{- \eps s} \, 1/s \, ds \ , \ \ \ \
 \V_D =  \frac{e^2 }{4 \pi} \sum _{i<j} \frac{v_i - v_j}{|v_i -v_j|^3} \, 
 (V_i - V_j)\, .$$

\vspace{8mm}


\end{document}